\documentclass[review]{elsarticle}

\usepackage{hyperref}
\usepackage{amsmath}
\usepackage{booktabs}

\journal{Neurocomputing}

\begin{document}

\begin{frontmatter}

\title{Time-Continuous Energy-Conservation Neural Network 
for Structural Dynamics Analysis }

\author[tusimple,ucdavis]{Yuan FENG\corref{mycorrespondingauthor}}

\cortext[mycorrespondingauthor]{Corresponding author}
\ead{yuanfeng.energy@gmail.com}

\author[ucdavis]{Hexiang WANG}
\author[ucdavis]{Han YANG}
\author[tianjin]{Fangbo WANG}

\address[tusimple]{TuSimple, 9191 Towne Centre Dr. STE 600. San Diego, CA 92122, United States}
\address[ucdavis]{University of California Davis, One Shields Ave, Davis, CA, 95616, United States}
\address[tianjin]{Tianjin University, No.92 Weijin Road, Nankai District, Tianjin, 300072, China}


\begin{abstract}
Fast and accurate structural dynamics analysis
is important for structural design and damage assessment.
Structural dynamics analysis leveraging 
machine learning techniques has become a popular 
research focus in recent years.
Although the basic neural network provides an alternative approach
for structural dynamics analysis, 
the lack of physics law inside the neural network
limits the model accuracy and fidelity.
In this paper, a new family of the energy-conservation neural network is introduced, which respects the physical laws.
The neural network is explored from a
fundamental single-degree-of-freedom system
to a complicated multiple-degrees-of-freedom system.
The damping force and external forces are also 
considered step by step.
To improve the parallelization of the algorithm,
the derivatives of the structural states are 
parameterized with the novel 
energy-conservation neural network
instead of specifying the 
discrete sequence of structural states.
The proposed model uses the system energy as the last layer 
of the neural network and leverages the underlying
automatic differentiation graph to 
incorporate the system energy naturally,
which ultimately improves the accuracy and long-term stability
of structures dynamics response calculation 
under an earthquake impact.
The trade-off between computation accuracy and speed is discussed.
As a case study, a 3-story building 
earthquake simulation is conducted 
with realistic earthquake records.
\end{abstract}


\end{frontmatter}


\section{Introduction}\label{sec_introduction}

With the explosive growth of data and 
cloud computing resources, 
recent advances in machine learning 
have yielded transformative results 
across diverse scientific disciplines, 
including 
computer vision~\cite{zhang2020},
natural language processing~\cite{devlin2018},
genomics~\cite{libbrecht2015},
and business intelligence~\cite{cai2017}
.
The main objective of machine learning is to grant the 
machine with strong generalization power 
using as few parameters as possible
~\cite{bishop2006pattern}.
The structural dynamics and physics law are not well-explored with
modern machine learning approaches.  
While there are some attempts to simulate the wave propagation using 
neural networks~\cite{hou2017new,kissas2019machine}
, it is unclear if the neural network follows the 
basic physics laws.
In addition, if the long sequence of discrete data is 
trained with neural network naively, 
the neural network becomes a very deep network,
which suffers from the vanishing or 
exploding gradient problems
~\cite{hanin2018neural,tan2019vanishing}.
Many tricks were studied to alleviate the problem, such as
non-saturating activation functions~\cite{chandar2019towards}, 
batch normalization~\cite{yuan2019generalized}, 
gradient clipping~\cite{kanai2017preventing}, 
and faster optimizers~\cite{kingma2014adam}.
Nevertheless, it is still tricky 
for a neural network to predict a long-term time-series
sequence.

The conventional approaches for structural dynamics
problems, like finite element methods, 
have the following difficulties. 

\begin{itemize}
\item Finite element analysis always involves
complicated modeling and meshing processes. 
The created mesh needs to capture the input domain
geometry with high-quality and well-shaped cells. 
In addition, the mesh should be adaptively refined in areas that 
are important for the subsequent calculations.
The complicate pre-processing and post-processing
procedure are cumbersome burdens for the users. 

\item %
Besides,  
when the material inelasticity or geometry nonlinearity plays a 
critical role in the dynamic response, 
the convergence of conventional approaches becomes an issue.
Many of the conventional numerical approaches 
require extensive 
computer simulations to achieve solid results.

\item
Last but not least, in finite elements analysis,
the discretization errors in the meshing and integration 
process are hard to control or estimate.
The finite element method usually gives the final results
anyway regardless of the mesh quality in the preprocessing steps. 
Therefore, users of the finite element analysis
must be a domain expert to 
justify the correctness and accuracy of the 
results~\cite{feng2019procedures}.
\end{itemize}

To avoid the problems above,
machine learning techniques, specifically neural networks, 
are explored to solve the structural dynamics problems
in this paper.
\begin{itemize}
\item 
Machine learning approaches operate on the 
data directly without the complicated 
geometrical preprocessing steps.

\item The activation function in the
neural network supports the nonlinearity 
natively, which can learn the 
material nonlinearity and geometrical nonlinearity
directly from the experimental data.

\item
The machine learning has a training 
dataset and a testing dataset, 
which lets the users control the error 
and the tolerance explicitly. 
The simplicity of straightforward learning
from the data democratizes the analysis
of structural dynamics problems.

\end{itemize}

While a real-world structure is complicated, this paper focuses
on the fundamental structural dynamics problem in single-degree-of-freedom (SDOF)
first and then 
generalizes to a multiple-degrees-of-freedom (MDOF) application.
The energy-conservation neural network is
designed to solve the structural dynamics problem 
in a physically informative way.
For any machine learning task, the quality, volume, and variety 
of the training data are crucial to the accuracy of the
prediction results.
In earthquake engineering, 
accessing to earthquake ground motion was hindered
by the large-scale of the data and the inconsistency
in how the data were gathered and stored.
Fortunately, nowadays the 
Pacific Earthquake Engineering Research (PEER) Center  
ground motion database~\cite{power2008}
and United States Geological Survey (USGS)
National Strong Motion Program~\cite{usgs2014}
provide large-scale ground motion records
for the next generalization of ground motion
prediction research.
The consistent and reliable worldwide ground motion records
built a cornerstone for this paper.
Specifically, the PEER strong motion database contains
over 174,000 records from 486 earthquake events 
and 1379 recording stations~\cite{usgs2014}.
The earthquake magnitude covers a 
range between 3.4 and 7.9, and a distance range between 0.05 and 1533 km. 
For each earthquake record, the available spectra are available for periods 0.01 to 20 seconds and the damping ratios ranges from 0.1 to 30\%.
Performance-based seismic design requires an
effective and efficient structural dynamics 
analysis approach.
Historically, the primary focus of building design
is to ensure life safety and collapse prevention.
Continued operation of the structure and 
reduction of economic losses associated with
earthquake damage to the structure 
are secondary considerations if they
are considered at all. 
Performance-based seismic design allows the 
engineer to choose the appropriate levels
of ground motion objectives and 
the level of protection for those ground motions.
Multiple levels of ground shaking 
are evaluated with a different level of 
performance specified for each level of 
ground shaking.
Nonlinear dynamic analysis of the structure
response plays a crucial role in seismic response prediction.
The proposed neural network in this paper
provides a novel approach to
estimate seismic demands at the operational performance level.
The details of the contributions in this paper are as follows.
\begin{itemize}
\item A versatile neural network framework with continuously-defined dynamics. 
The framework is able to naturally 
exploit observed data that arrives at arbitrary times.
After the training, it is also 
convenient to predict model states at any 
arbitrary times. 
In addition, the prediction to the derivative can be 
computed in parallel, which has a
better scalability.

\item A novel energy conservation neural network that
incorporate the system energy natively.
The proposed model is physically informed,
which predict the structural dynamics response
realistically.

\item An efficient machine learning approach
for structural dynamics problems 
from a fundamental single-degree-of-freedom system
to a complicated multiple-degrees-of-freedom system.
The proposed model potentially opens the path
for the data-driven computation of structural dynamics problems.

\end{itemize}

\section{Background and Related Work}\label{sec_related_work}

Among various machine learning techniques, the conventional
neural network for predicting time series is recurrent neural network (RNN)
~\cite{rumelhart1986}.
RNNs process a time series step-by-step, maintaining an internal state summarizing the information
they have seen so far.
Basically, the output of the neuron is feed as a part of the input to the same neuron again and again.
To train an RNN on long sequences, the neuron needs to run over many time steps, 
making the unrolled RNN a very deep network,
which may have prohibitive time complexity 
in practical situations.
%
%
%
%
%
To fix the issues in RNNs, various types of cells with long-term
memory have been introduced. 
The popular types are long short term memory (LSTM) 
~\cite{hochreiter1997} 
cell, gated recurrent unit (GRU)
~\cite{cho2014}
cell, and the attention mechanism 
in the Transformer model~\cite{devlin2018bert}.
%
Transformer uses 12 separate attention mechanism 
without using sequence aligned convolution, which is convenient 
for the large-scale parallelism on GPU.
Although Transformer achieves state of the art performance in natural language processing (NLP),
the time series data won't have exactly the same events in NLP. 
And since the time series data lack repeated tokens, 
Transformer may not be applied to time series prediction directly.


Besides the conventional neural network for time series forecasting, 
customized neural network have been designed and applied 
to solve structural dynamics and wave propagation problems.
These works are classified into four categories.

The first category of works focus on wave physics.
For example, 
Hughes~\cite{Hughes2019} 
built a mapping between the dynamics of wave physics and the computation
in the recurrent neural network. 
The mapping leveraged physical wave systems to learn complex features
in temporal data, which is in the opposite direction of this work.
Zhu~\cite{Zhu2017,zhu2019phasenet} applied the neural network to predict the propagation of wavefront 
which contains complex wave-dynamics phenomena like velocity anomalies,
reflection, and diffraction.
Sorteberg~\cite{Sorteberg2019} built a predictive neural network comprising of autoencoder
and LSTM (long short term memory) cells. The neural network 
can predict at most 80 time-steps of wave propagation.

The second category of works relies on a large-scale of observed data to extract the dynamics features.
For instance,
Mousavi~\cite{mousavi2019cred} trained the CNN-RNN earthquake detector based on
deep neural network. 
A large-scale of seismograms were feed into the neural network to detect the earthquake signals.
Broggini~\cite{broggini2014data} proposed a data-driven wave-field focusing which retrieves the 
Green's function recorded at the acquisition surface due to a virtual source located at depth.
Besides, Cao~\cite{cao2013seismic} applied a seismic data-driven rock physics model to estimate 
the partial saturation effects in tight rocks~\cite{feng2014current}. 

The third category of works explores the relation between neural network and wave differential equation.
Lu~\cite{lu2017beyond} proposed a new linear multi-step (LM) architecture 
inspired by the linear multi-step method solving ordinary differential equations.
The new architecture was used to achieve higher accuracy on image classification.
Zhu~\cite{zhu2018convolutional} introduced Runge-Kutta method to build a convolutional neural network 
and achieved superior accuracy. 
Chen~\cite{chen2018neural} demonstrated the neural ordinary differential equations 
in continuous-depth residual networks and continuous-time latent variable models.

The fourth category of works applied end-to-end machine learning models to
specific applications.
Bagriacik~\cite{bagriacik2018} investigated the earthquake damage to water pipelines 
using statistical and machine learning approaches.
Kiani~\cite{kiani2019} applied machine learning techniques 
for deriving seismic fragility curves.
Song~\cite{song2018} predicted structural responses accurately and 
identified salient attributes using the generalized additive model.

While the existing researches focus on wave propagation,
signal detection, and differential equations.
This research focuses on the structural dynamics problem in earthquake engineering.
This paper built a novel time-continuous energy-conservation neural network 
for structural dynamics analysis.


\section{Baseline Time-Continuous Neural Network}\label{sec_basenn}

Instead of training the discrete signal directly using RNN, 
the neural network in this paper 
parameterizes the \textit{derivative of structural states} 
with respect to time.
For a classic single-degree-of-freedom (SDOF) mass-spring-damper system, 
the equilibrium equation using Newton's second law of motion is 
\begin{equation}
	m\ddot{u} + c\dot{u} + f_S(u,\dot{u}) = p(t)
	\label{eq_equiv_newton2}
\end{equation}
where m is the mass, $u$ is the displacement, 
$c$ is the viscous damping coefficient,
$f_S$ is the elastic or inelastic resisting force from the structure,
and $p(t)$ is the external force.
In Equation~\ref{eq_equiv_newton2},
$m\ddot{u} $ is the inertial force
and $c\dot{u} $ represents the damping force.
The left side of Equation~\ref{eq_equiv_newton2}
is the summation of all the internal forces,
which is equal to the right side,
the external force $p(t)$.

Instead of building a neural network solving the equation above,
we describe the motion in an incremental perspective

\begin{equation}
	u_{t+1} = u_{t} + f ( u_{t}, \theta_{t})
\end{equation}
where $\theta_t$ is the state variable at step $t$,
and $f(u_t, \theta_t)$ is the displacement increment.

The dynamics can be parameterized as an ordinary 
differential equation (ODE) specified by a neural network:
\begin{equation}
	\frac{d\mathbf{u}(t)}{dt} = f ( \mathbf{u}(t), \theta)
	\label{eq_continuous_base_nn}
\end{equation}

\begin{figure}
  \centering
  \includegraphics[width = 9cm]{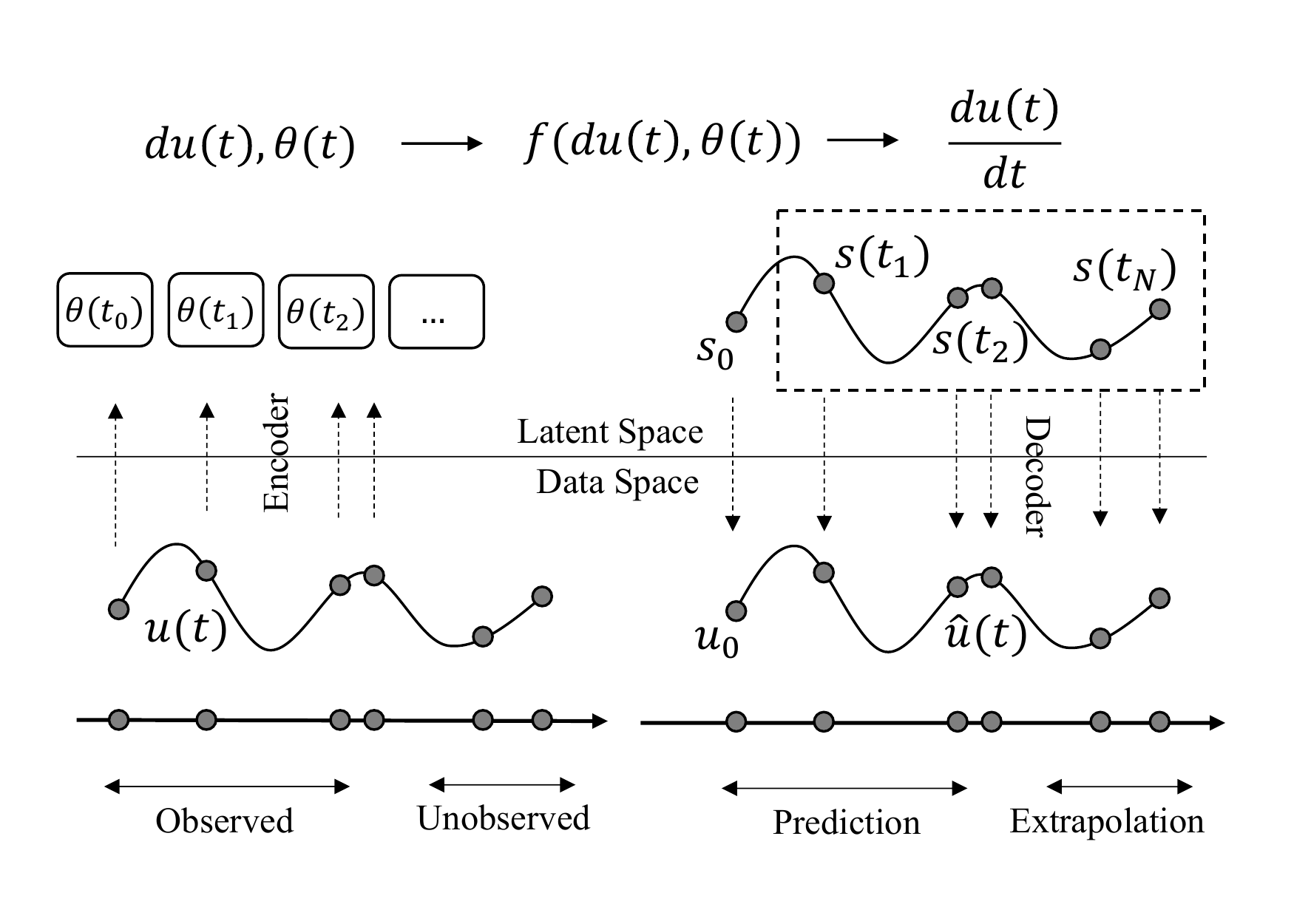}
  \caption{Computation graph of neural network. }
  \label{fig_computation_graph_of_neural_network}
\end{figure}

The neural network in this paper focuses on the approximation of
the time-continuous function $f ( \mathbf{u}(t), \theta)$ 
in Equation~\ref{eq_continuous_base_nn}.
The computation graph of the neural network is shown in 
Figure~\ref{fig_computation_graph_of_neural_network}.

A classic neural network is trained to predict
the time series derivative from the initial state of the SDOF problem.
The initial state $\mathbf{u}(0)$ is considered as the input layer $\mathbf{u}(0)$.
The following states $\mathbf{u}(T)$ is defined as the output layers,
which can be computed by the neural network.
Defining and evaluating the time series model using differential equation has
the following benefits:

\textbf{Continuous time-series models}
In contrast to RNNs, which require discrete observed data in fixed time steps, 
continuously-defined dynamics can naturally incorporate data that arrives
at arbitrary times. This method makes it easier to collect observation data.
After the training, the method is also easier to predict model states
at an arbitrary time.

\textbf{Scalable Inference}
While the time-continuous equation must be integrated sequentially, the inference
from the displacement to the derivative can be computed in parallel.
This feature ensures the accessibility of this method for a large-scale 
dynamics system.

\section{Theory of proposed Neural Network}\label{sec_theory}

On top of the baseline time-continuous neural network in the previous section, 
the energy conservation neural network is proposed. 
The idea of the proposed neural network is inspired
by Hamiltonian mechanics~\cite{devlin2018bert,greydanus2019hamiltonian}.

\textbf{Hamiltonian Mechanics}

Hamiltonian mechanics describes the Lagrangian mechanics in a different set 
of mathematical formulation. 
In Hamiltonian mechanics, the physical system is described in 
canonical coordinates $\mathbf{r = (q,p)} $.
The vector $\mathbf{q}$ are the generalized coordinates,
which represents the positions of the targeting objects.
The vector $\mathbf{p}$ denotes the corresponding momentum.
The time evolution of the system is defined by Hamilton's equations:
\begin{equation}
	\frac{d\mathbf{p}}{dt} = - \frac{\partial H}{\partial \mathbf{q}},
	\frac{d\mathbf{q}}{dt} = + \frac{\partial H}{\partial \mathbf{p}}
	\label{eq_hamilton}
\end{equation}
where $ H = H(q,p,t) $ is the Hamiltonian, which 
corresponds to the total energy of the system in this paper.
%
The Hamiltonian equation establishes the framework
to relate the position and the moment vectors
$(p,q)$ to the total energy.
So the motivation to use Hamiltonian mechanics 
is to describe the vibration system 
in which energy changes from kinetic to potential 
and back again over time.

\textbf{Energy Conservation Neural Network}

In this paper, an energy conservation neural network (ECNN) is proposed
to learn the structural dynamics function.
The network is able to learn exactly conserved quantities from data. 
During the forward pass, the model consumes the displacement
and velocity and outputs the energy.
Next, the in-graph gradient of the energy with respect 
to the input coordinates are automatically computed.
Then, with respect to the gradient, the $L_{2} $ loss 
is computed and optimized.
\begin{equation}
	\mathcal{L}_{\text {ECNN}} = 
	    \left\| \frac{d\mathbf{p}}{dt} + \frac{\partial E}{\partial \mathbf{q}} \right\|_2
	  + \left\| \frac{d\mathbf{q}}{dt} - \frac{\partial E}{\partial \mathbf{p}} \right\|_2
	\label{eq_loss_energy_ecnn}
\end{equation}

\begin{figure}
  \centering
  \includegraphics[width = 7cm]{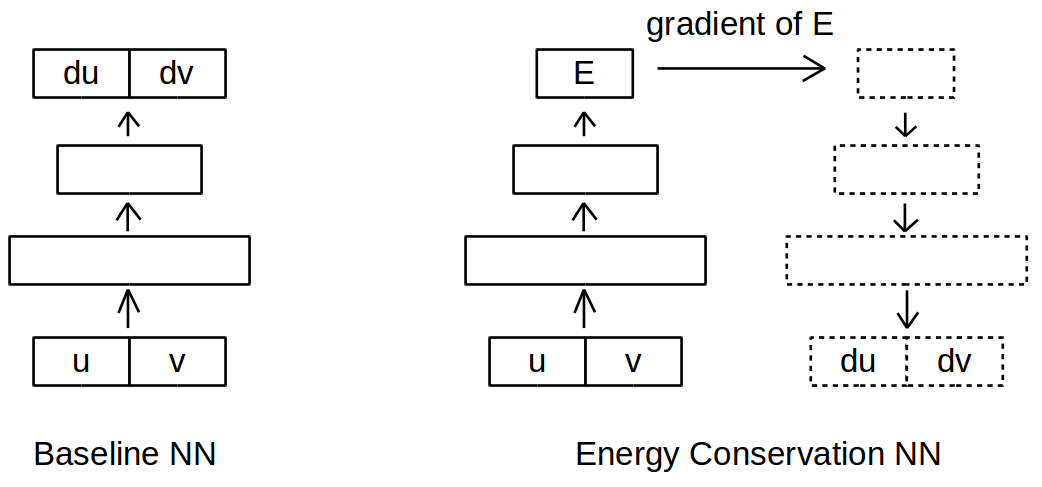}
  \caption{Architecture comparison between baseline neural network
  and energy conservation neural network. 
  In the figure, the blocks represent layers of neural network (NN). 
  $u$ and $v$ represents the earthquake wave 
  displacement and velocity respectively. 
  $du$ and $dv$ represent the derivative of 
  $u$ and $v$ with respect to time.}
  \label{fig_base_nn_vs_ecnn_arch}
\end{figure}

The architecture comparison between the baseline neural network
and ECNN is shown 
in Figure~\ref{fig_base_nn_vs_ecnn_arch}.

For an undamped free vibration, the energy input to an SDF system 
is controlled by the initial displacement $u(0)$ and 
initial velocity $\dot{u}(0) $.
At any instant of time, the total energy in a freely vibrating 
system is made up of two parts, kinetic energy $E_{\text K} $ of the mass
and potential energy equal to the strain energy $E_{\text S} $
of deformation in the spring.  
Considering the input energy in an undamped system,
the total energy is 
\begin{equation}
	E_{\text K}(t) + E_{\text S}(t) = E_{\text {tot}}(t) = \frac{1}{2}k[u(0)]^2 + \frac{1}{2}m[\dot{u}(0)]^2
	\label{eq_energy_conservation_nn}
\end{equation}

Equation~\ref{eq_energy_conservation_nn} tells us that 
the displacement and velocity in structural dynamics keep 
the total energy exactly constant over time.
The mathematical framework relates the displacement and velocity
of a system to its total energy $E_{\text {tot}} = G(u, \dot{u}) $.
This is a powerful approach because it engraves
the total energy to the neural network.
The model can predict the motion in an SDOF system, 
but its true strengths 
only become apparent 
when tackling systems with multiple degrees of freedom (MDOF).

\begin{figure}
  \centering
  \includegraphics[width = 6cm]{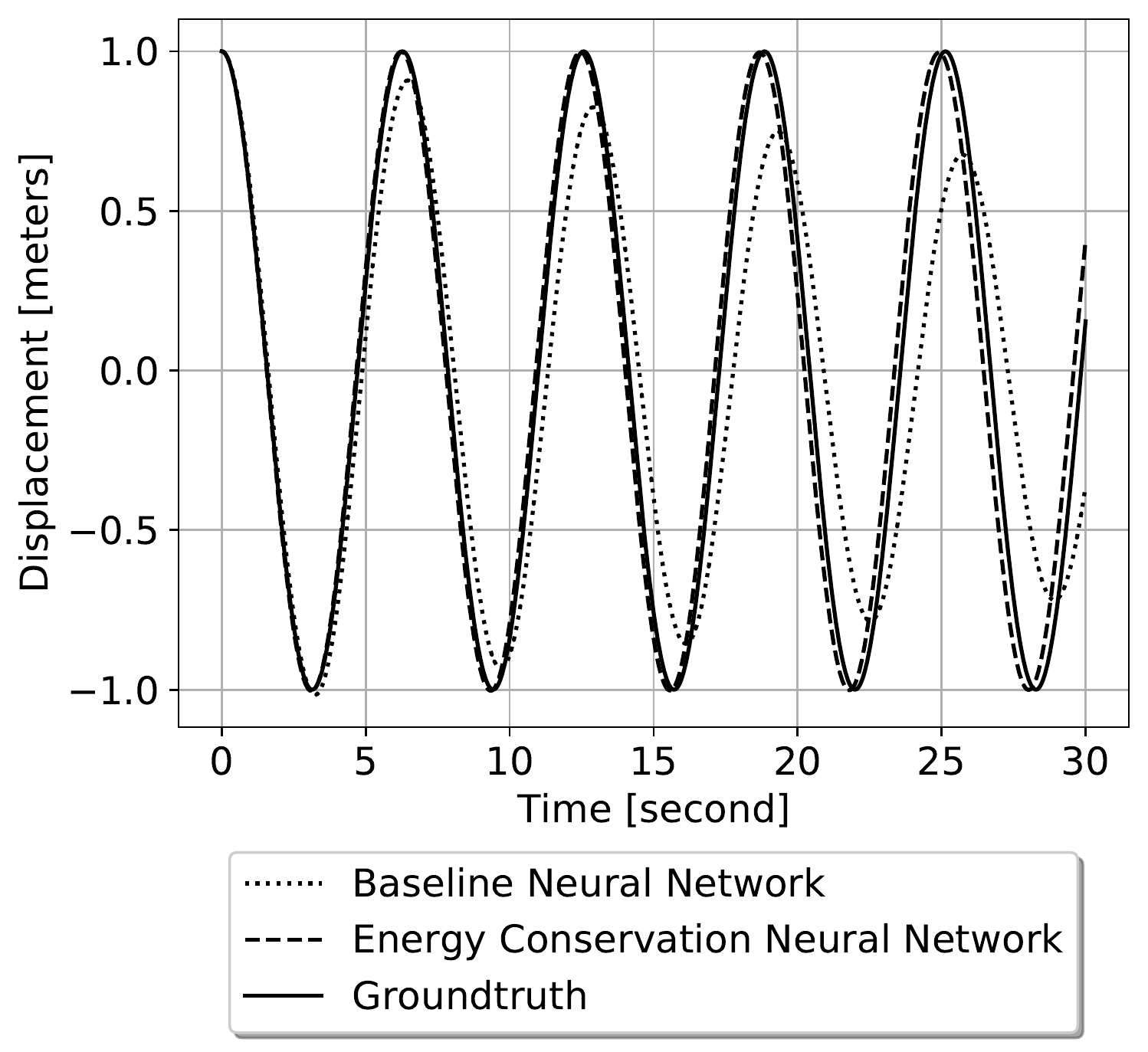}
  \includegraphics[width = 6cm]{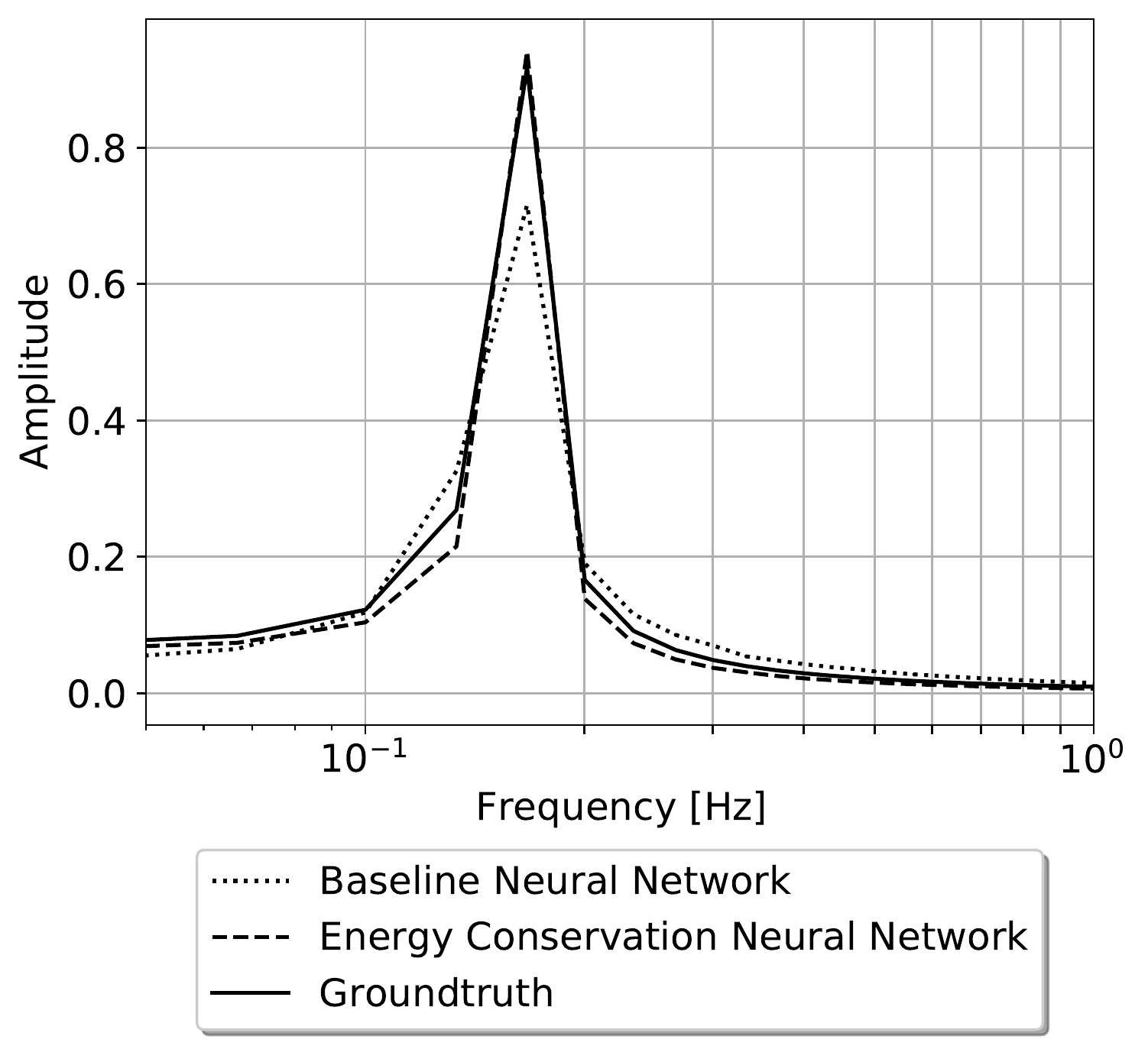}
  \includegraphics[width = 6cm]{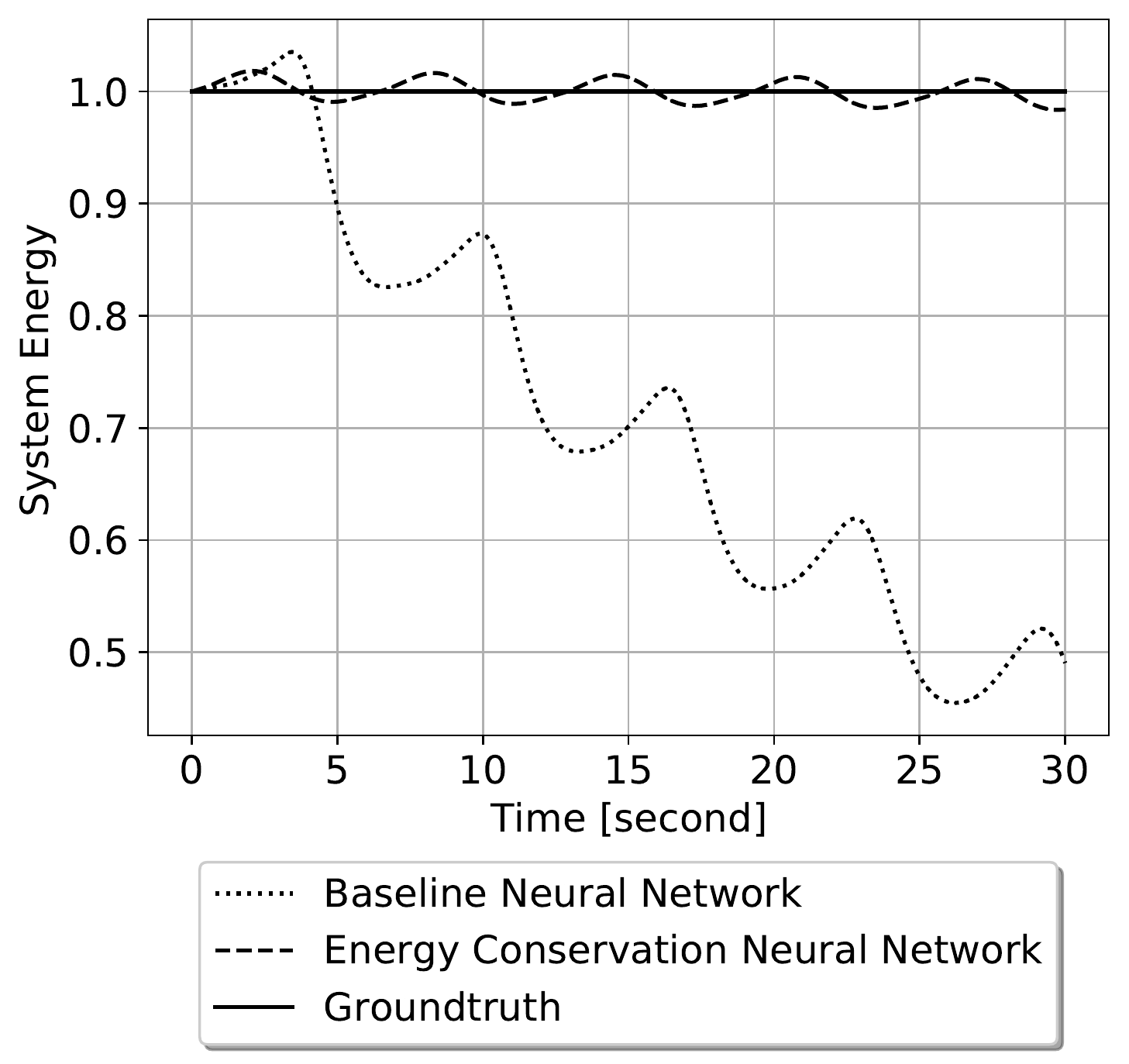}
  \caption{Comparison of displacement and system energy 
  between baseline neural network and energy-constant neural network. }
  \label{fig_sdof_constant_comparison}
\end{figure}

The illustrative comparison result between ECNN 
and 
the baseline neural network 
is shown in Figure~\ref{fig_sdof_constant_comparison}
and Figure~\ref{fig_sdof_damping_comparison}.
The procedures in ECNN allow the conserved quantities of 
total energy to be learned from the data.
With the physics laws, ECNN establishes the perfect mapping from
the dynamics states to their corresponding derivatives.
Besides, the conservation law can be manipulated 
to control the total energy
in the structural dynamics system.

\section{Single Degree of Freedom}\label{sec_sdof}
\subsection{Damping Effects}\label{subsec_damping}
Besides the energy conservation properties, in this section,
we explore if the energy conservation properties can be 
manipulated to simulate the damping effects in a structural dynamics system.
The exact damping effects are 
hard to be quantized in conventional structural dynamics
~\cite{sinha20173}.
The reasons are as follows.

First, the damping matrix cannot be determined by individual structural elements
like the stiffness matrix because the damping properties 
of various construction materials
are usually not well defined.

Secondly, a significant part of the damping effect is from the energy
dissipated in the opening and closing of micro-cracks in 
concrete~\cite{wang2018characterization},
friction at steel connections, and nonstructural elements, like
heavy equipment and indoor decoration.
Traditionally, classical damping, like Rayleigh or Caughey damping,
are used when there are similar damping mechanisms
distributed throughout the structure
~\cite{yang2017study}.
However, when the dynamics system consists of multiple parts with
different levels of damping, 
classical damping is no longer 
appropriate. 
For example, in a soil-structural interaction system, 
the damping of the structural part
is usually only one-tenth of the soil part.

\begin{figure}
  \centering
  \includegraphics[width = 6cm]{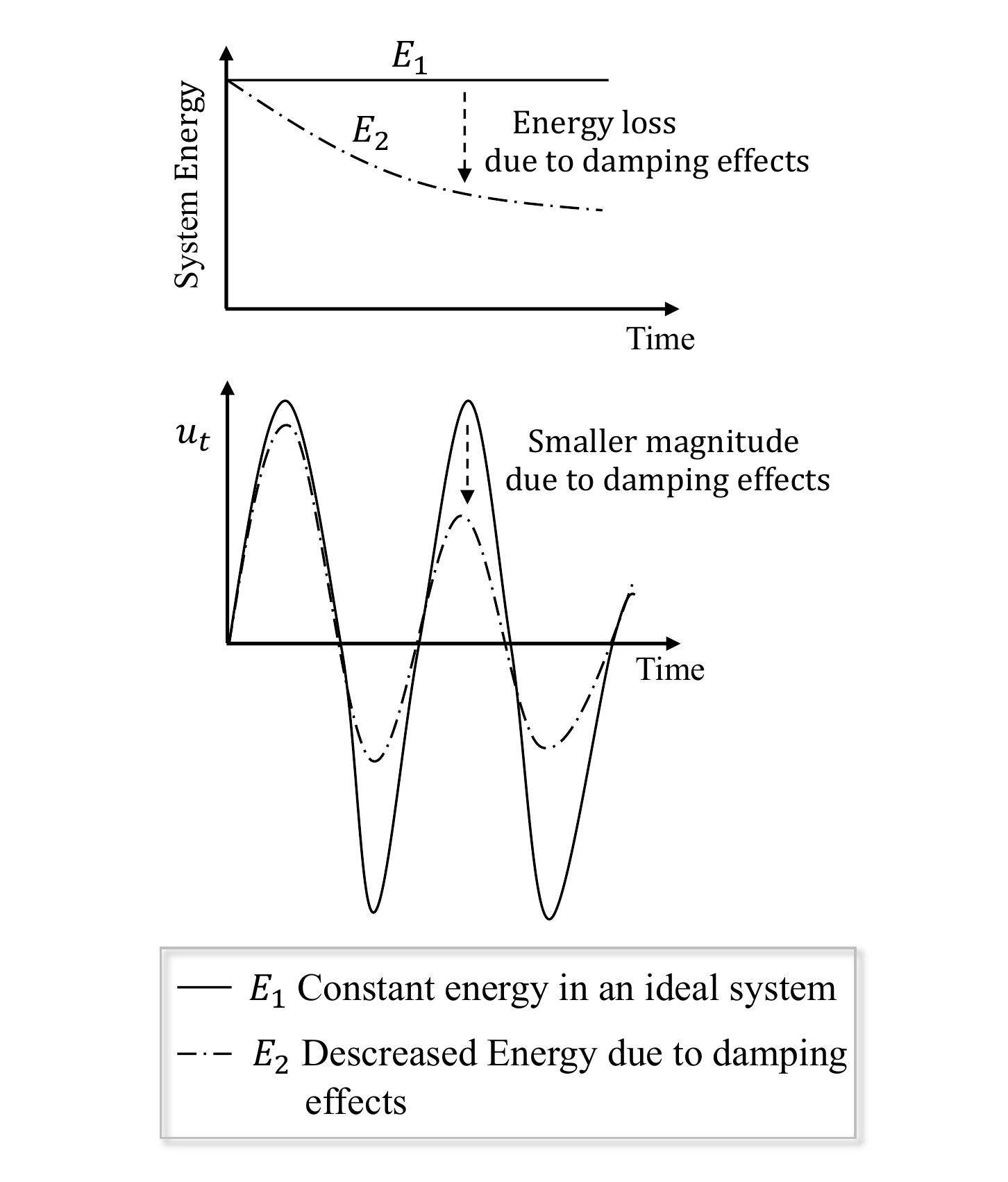}
  \caption{Damping effects in energy conservation neural network.
    In the figure, the system energy decreases with time due to the 
    damping effects. In addition, the maximum displacement magnitude $u_t$ 
    also decreases with time.}
  \label{fig_apply_damping_effects_ecnn}
\end{figure}

ECNN provides another approach to quantize the damping effects.
The damping effects are applied by controlling
the system energy in ECNN, as 
shown in Figure~\ref{fig_apply_damping_effects_ecnn}.
For each node, the predicted derivative of displacement is 
modified in Equation~\ref{eq_damping_effect}
with the consideration of damping effects.
\begin{equation}
	\frac{du}{dt} = e^{-\xi t} (\frac{du}{dt})_{\text {original}}
	\label{eq_damping_effect}
\end{equation}
where $\xi$ represents the damping effect and 
a bigger $\xi$ represents a larger damping effect.

\begin{figure}
  \centering
  \includegraphics[width = 6cm]{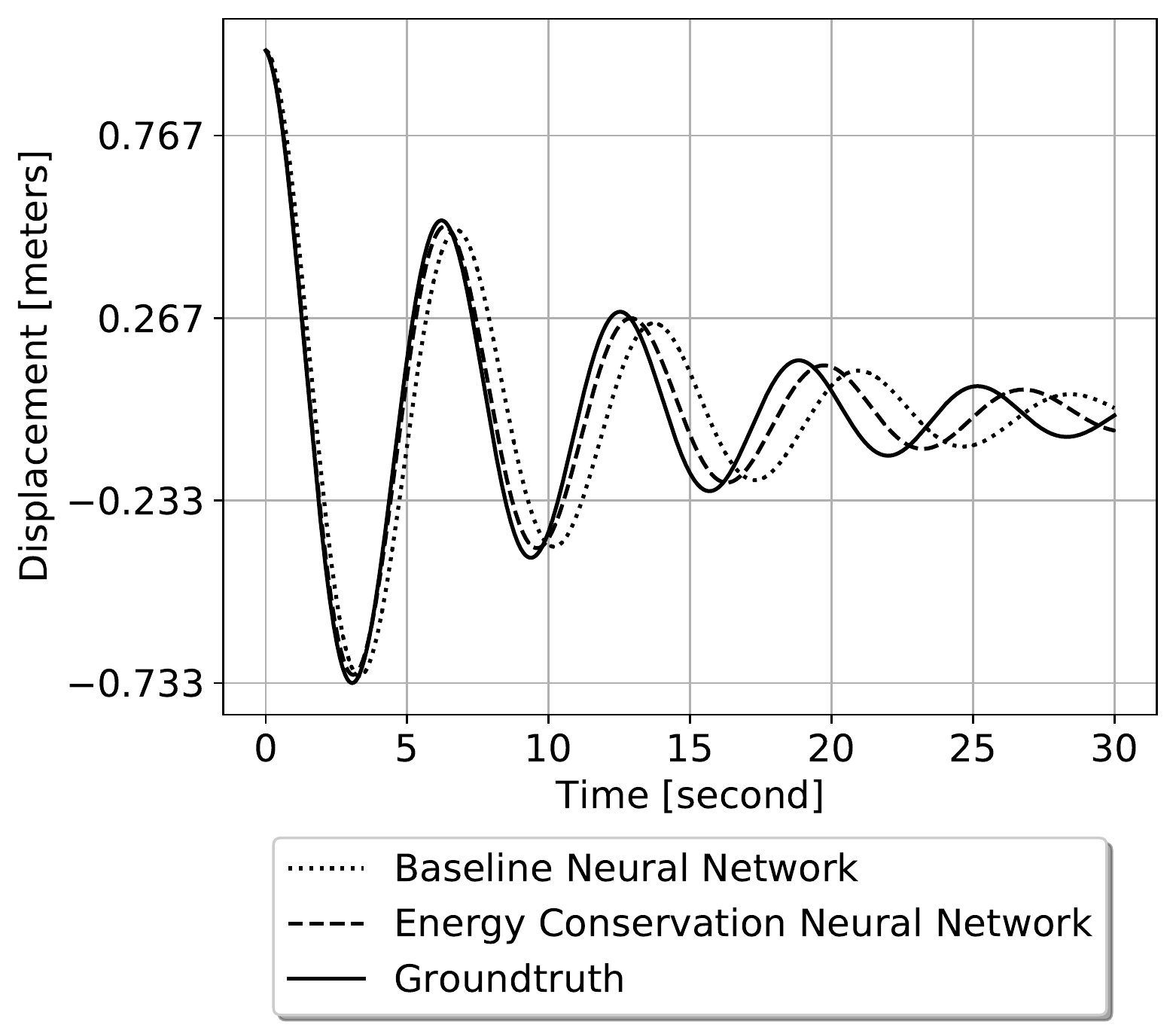}
  \includegraphics[width = 6cm]{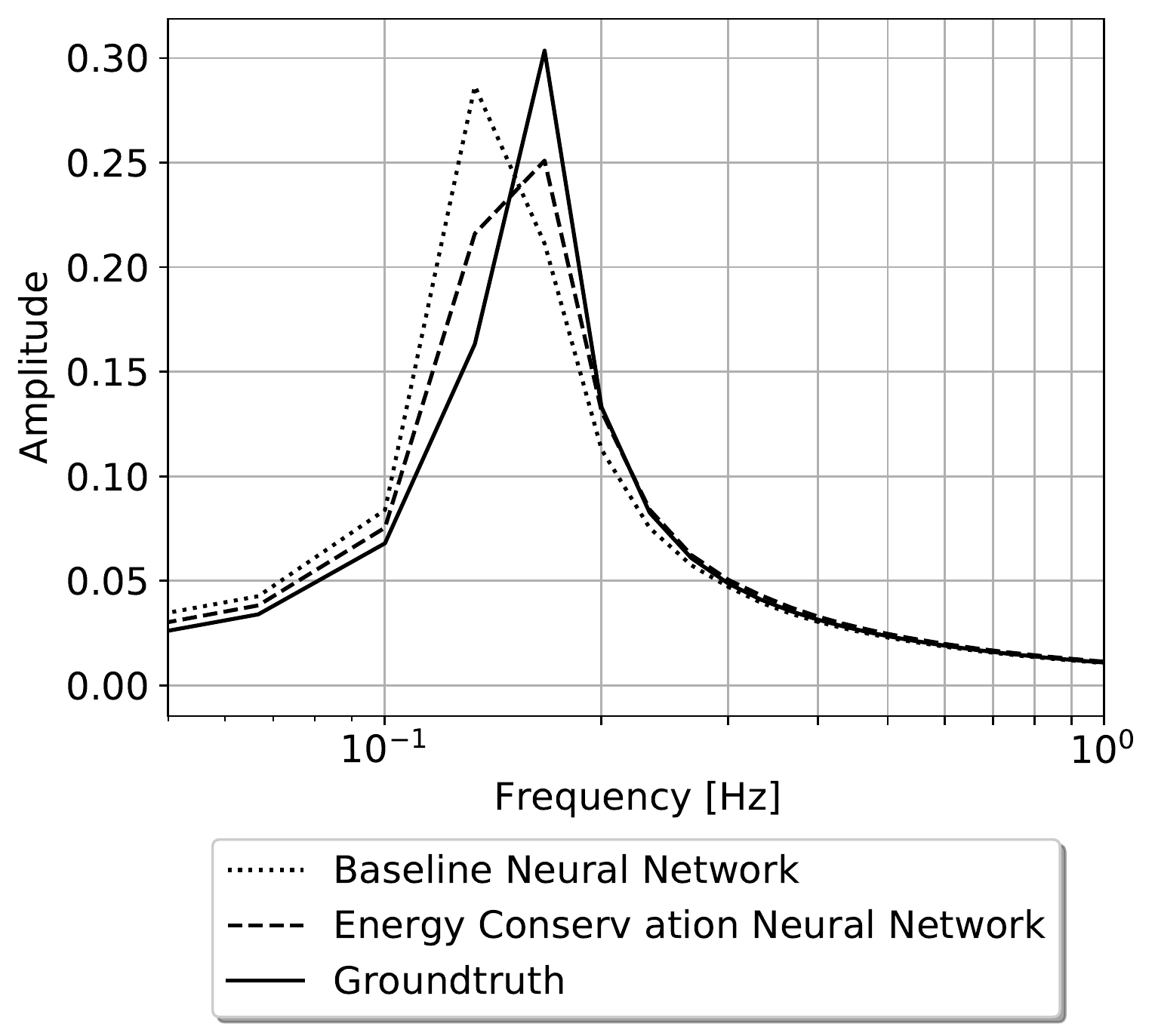}
  \includegraphics[width = 6cm]{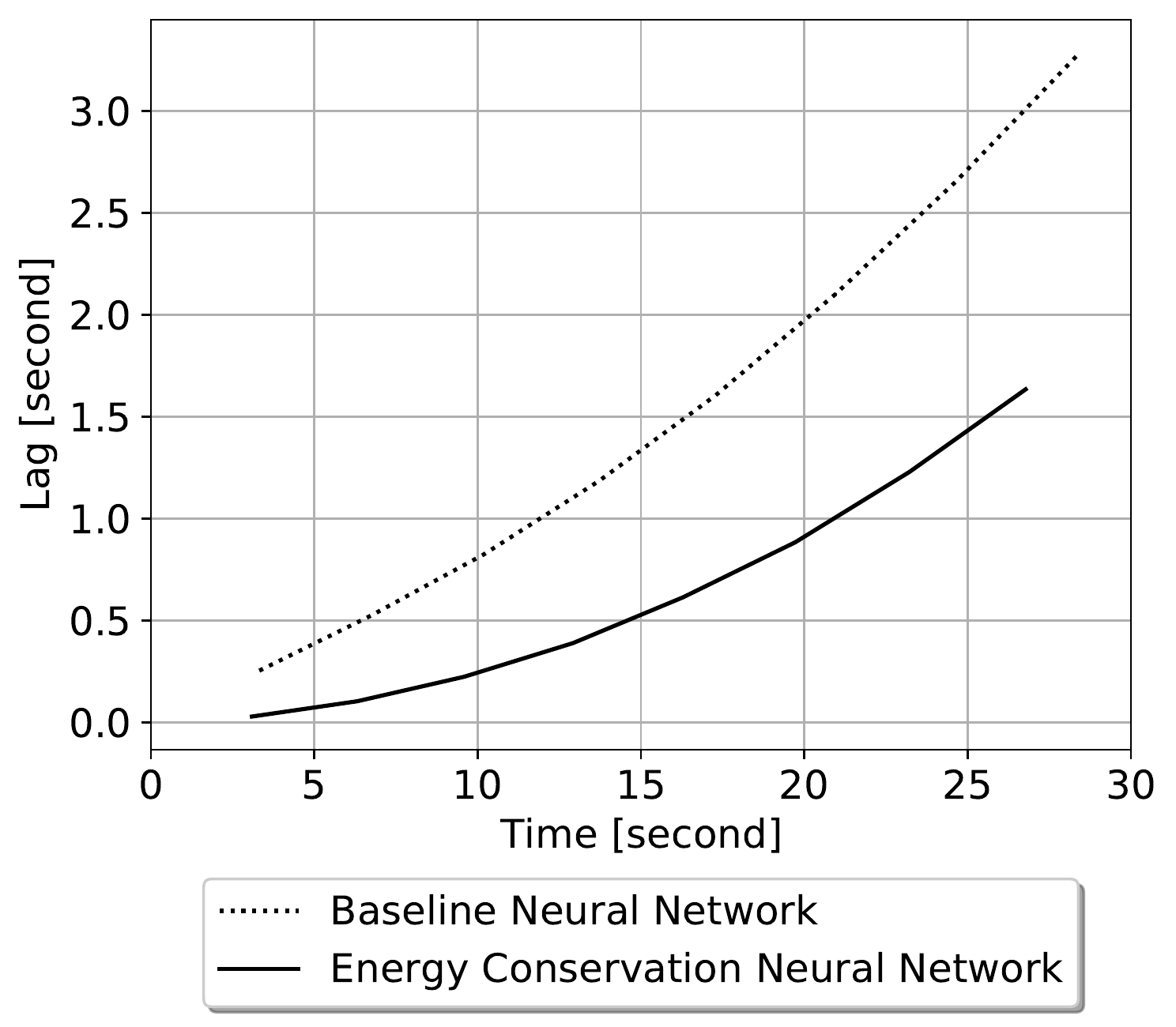}
  \caption{Comparison of displacement and phase lag
  between baseline neural network and energy-constant neural network. }
  \label{fig_sdof_damping_comparison}
\end{figure}

As shown in Figure~\ref{fig_sdof_damping_comparison},
ECNN has 50\% less phase lag error in damping
effects calculation
compared to the baseline neural network.

\subsection{External Forces}\label{subsec_externalforce}
While the damping force dissipates energy, 
the external force injects energy to the system.
The energy dissipated by viscous damping in one cycle 
of harmonic vibration is 
\begin{equation}
\begin{split}
E_{\text D} & = \int f_{\text D} du = \int_0^{2\pi/\omega} (c \dot{u}) \dot{u}\ dt = \int_0^{2\pi/\omega} c \dot{u}^2 \ dt \\
    & = c \int_0^{2\pi/\omega} [\omega u_0 \cos(\omega t - \phi)]^2 dt = \pi c \omega u_0^2 = 2\pi \xi \frac{\omega}{\omega_n} k u_0^2
\end{split}
\end{equation}

The energy dissipated is proportional to the square of the amplitude of motion. 
The energy dissipated increases linearly with excitation frequency.
On the other side, an external force $p(t)$ injects energy to the system, 
which is 
\begin{equation}
\begin{split}
E_{\text I} & = \int p(t)\ du = \int_0^{2\pi/\omega} p(t) \dot{u} \ dt \\
	& =  \int_0^{2\pi/\omega} [p_0 \sin\omega t] [ \omega u_0 \cos(\omega t - \phi)] dt = \pi p_0 u_0 \sin\phi
\end{split}
\end{equation}
for each cycle of vibration.

\begin{figure}
  \centering
  \includegraphics[width = 5cm]{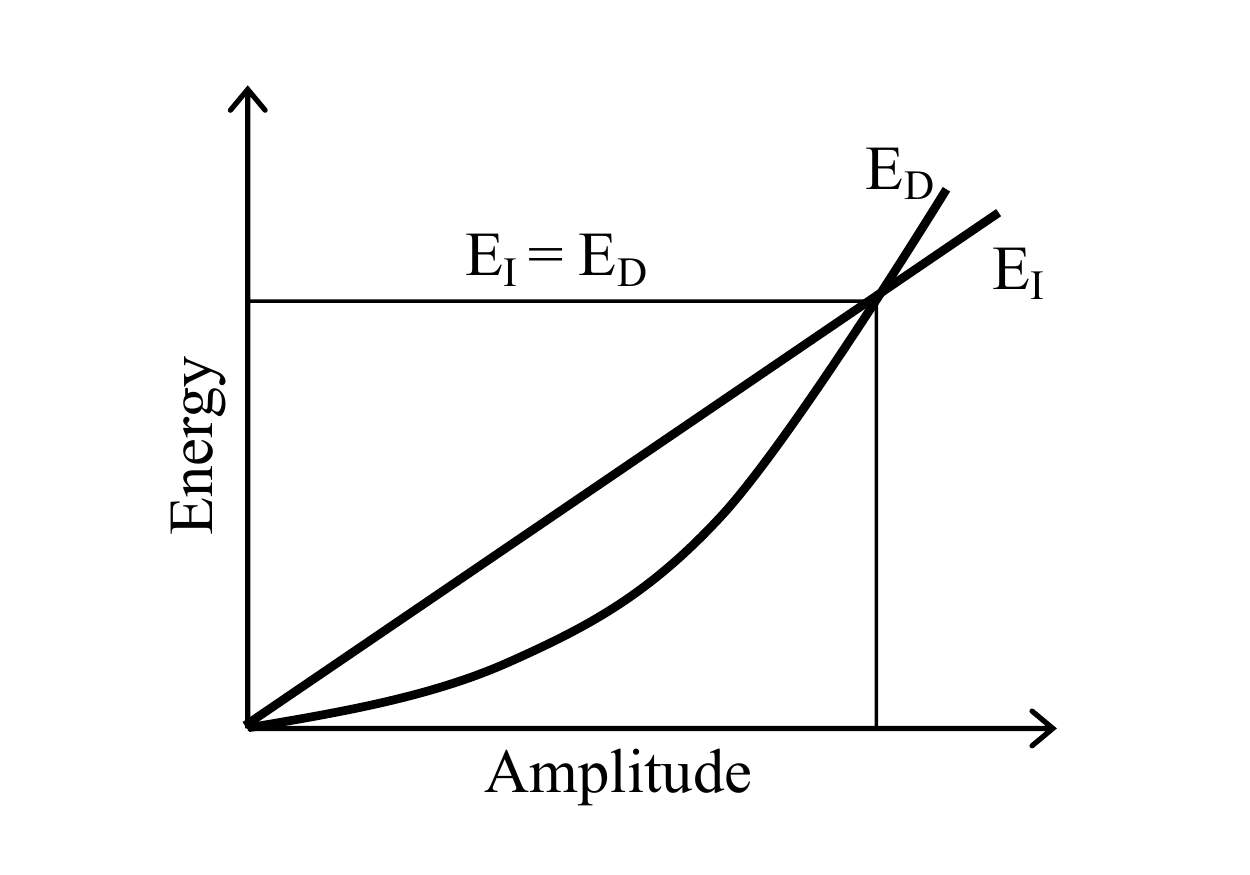}
  \caption{The relationship between the input energy $E_{\text I} $
    and the damping energy $E_{\text D}$ in a dynamic system. 
    The system energy becomes stable when the input energy
    $E_{\text I}$ equals to the damping energy $E_{\text D}$. }
  \label{fig_energy_input_dissipated_equal}
\end{figure}

In a steady-state vibration, the energy input to the system matches 
dissipated energy in viscous damping, namely $E_{\text D}=E_{\text I}$, which gives 
\begin{equation}
u_0 = \frac{p_0}{c \omega_n}
\end{equation}
The relation between the input energy and the dissipated energy 
is plotted in 
Figure~\ref{fig_energy_input_dissipated_equal}.

Leveraging the constant energy condition in this steady-state, 
the ground-truth motion is as follows.
\begin{equation}
\begin{split}
u_t & = \frac{p_0}{k}\frac{1-(\omega/\omega_n)^2}{[1-(\omega/\omega_n)^2]^2 + [2\zeta(\omega/\omega_n)]^2} \sin\omega t \\
	& + \frac{p_0}{k}\frac{-2\zeta \omega/\omega_n}{[1-(\omega/\omega_n)^2]^2 + [2\zeta(\omega/\omega_n)]^2} \cos\omega t
\end{split}
\end{equation}

where $p_0$ is the maximum value of the periodic force,
whose frequency $\omega$ is called exciting frequency.
Frequency $\omega_n$ is the natural frequency of the SDOF system,
and $k$ represents the stiffness of the system,
$\zeta$ represents the viscous damping of the system.

\begin{figure}
  \centering
  \includegraphics[width = 6cm]{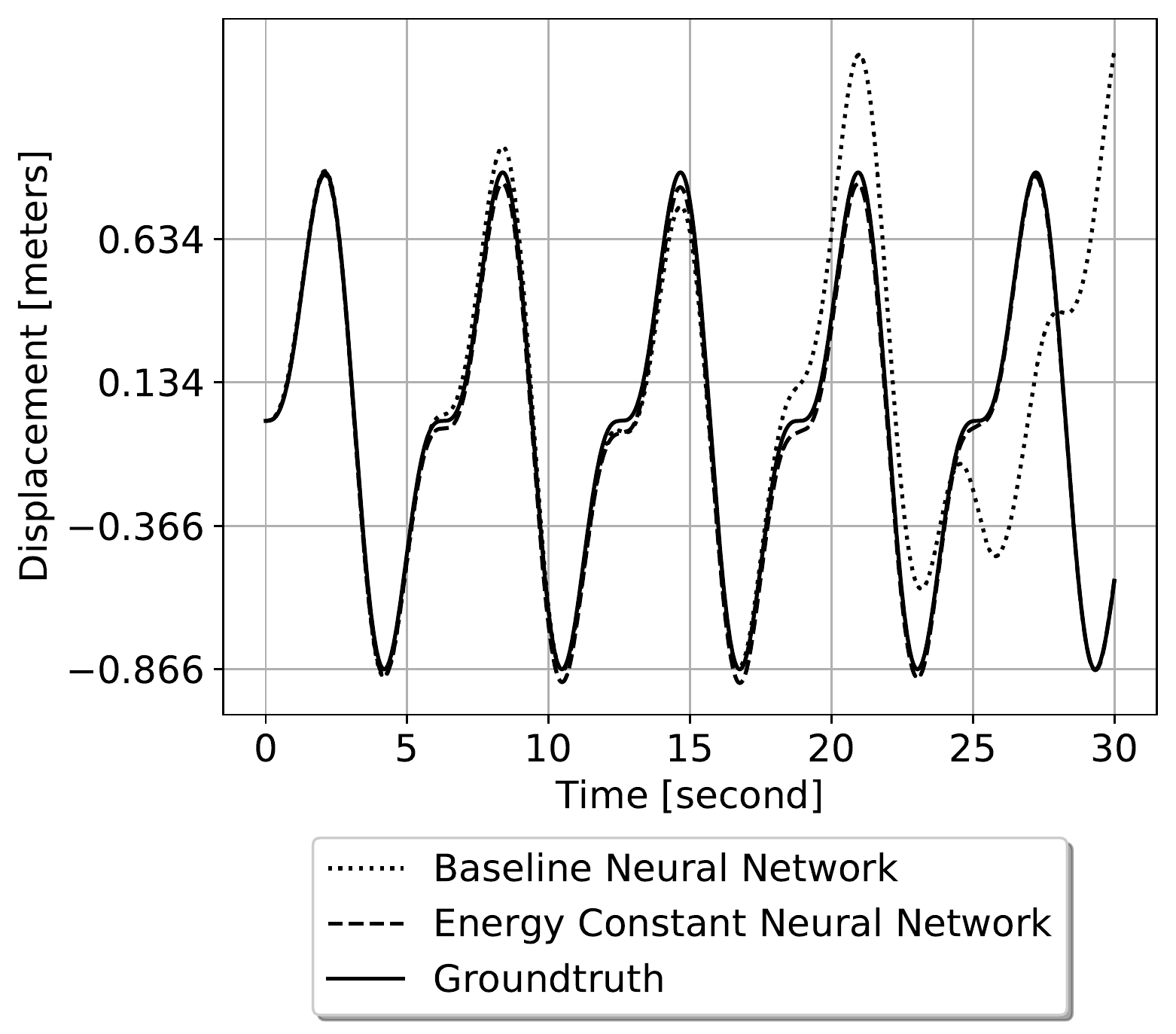}
  \includegraphics[width = 6cm]{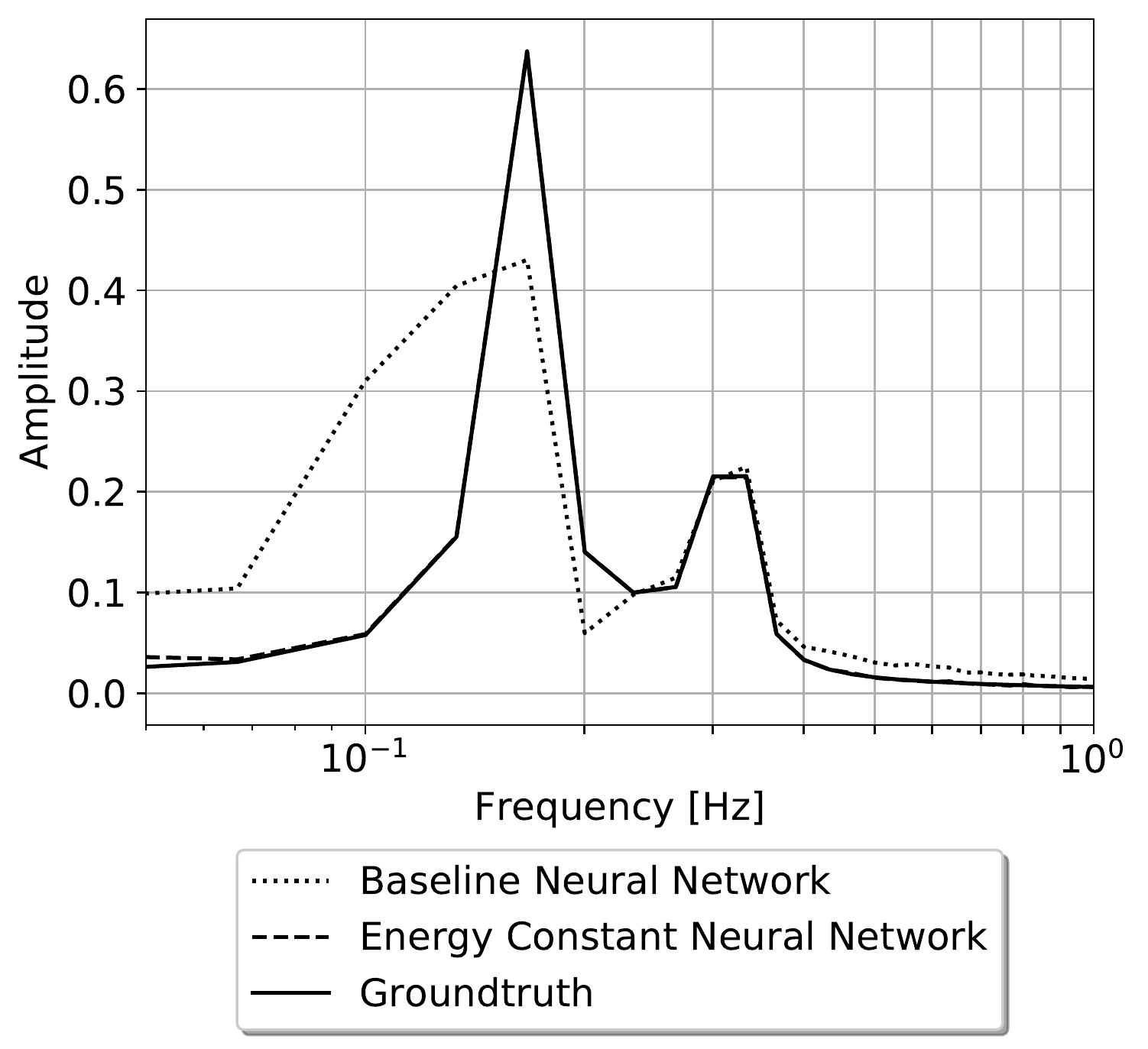}
  \includegraphics[width = 6cm]{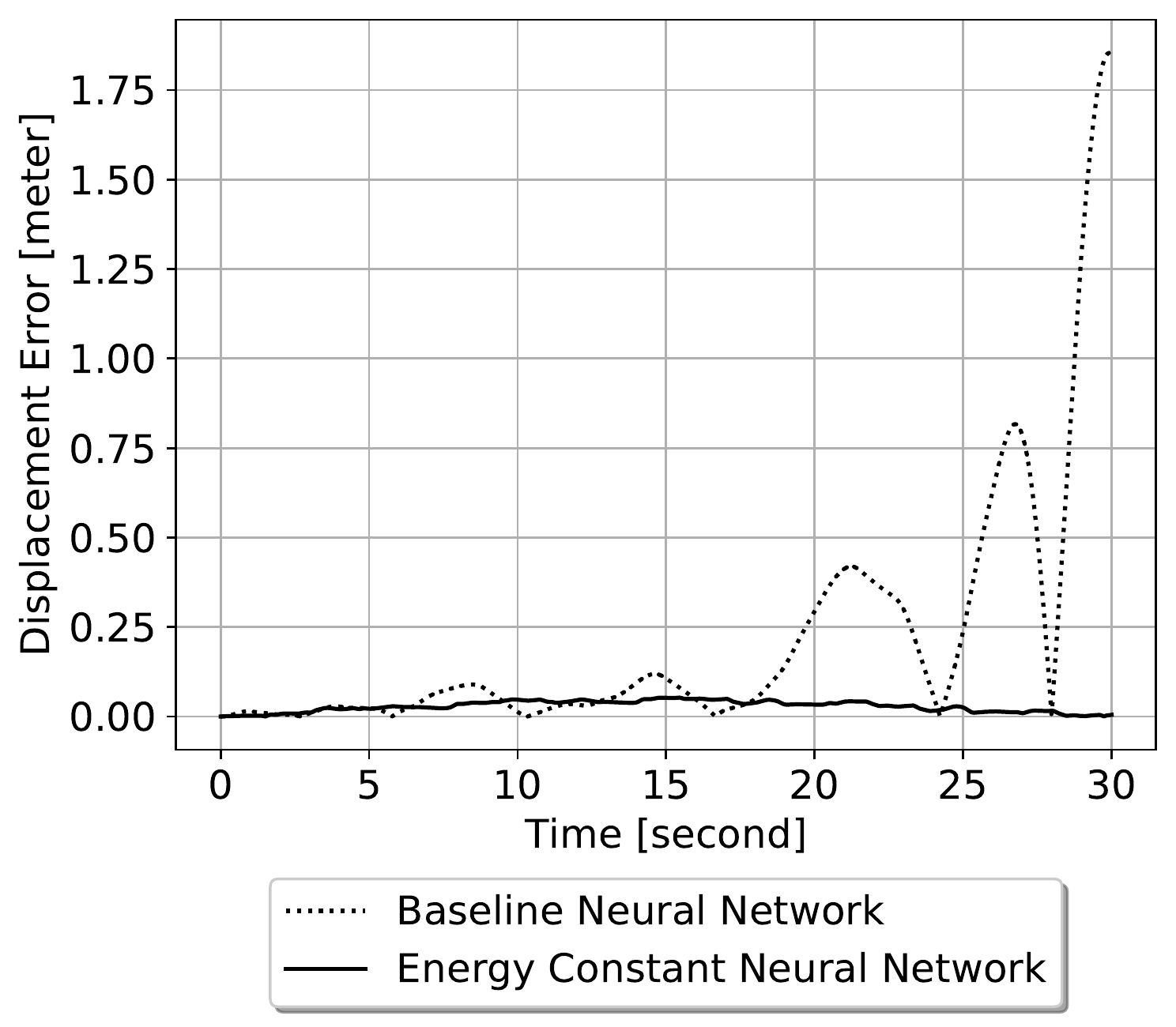}
  \caption{Comparison of displacement and error 
  under a periodic external force
  between baseline neural network and energy-constant neural network. }
  \label{fig_sdof_external_force_error}
\end{figure}

In numerical experiments, 
the predicted displacements using 
the baseline model and ECNN are 
shown in Figure~\ref{fig_sdof_external_force_error}.
The baseline model results drift away
after 3 cycles of prediction,
but the ECNN model prediction stays stable
across the entire experiment.

\section{Multiple Degrees of Freedom}\label{sec_mdof}

In previous sections, structural dynamics problem
is formulated for structures discretized as an SDOF system.
However, in real-world applications, 
most buildings and structures
are actually MDOF systems.
Generalization of the neural network from the SDOF system to
MDOF system is crucial for real-world applications.

In an MDOF system, structure states are composed of 
dynamics responses at all components.
These responses contain both essential and redundant information
about the system.
To reduce the computation complexity, 
a selection of important targets is predicted 
using the baseline neural network 
and energy conservation neural network.
The objective function of neural networks is
in Equation~\ref{eq_mdof_objective_function}.
\begin{equation}
u_{\text{target}}(\omega) = g(u_{\text{source}}(\omega), \theta(\omega))
\label{eq_mdof_objective_function}
\end{equation}
where $\omega$ is the frequency of the motion,
$\theta$ represents the structure states,
$u_{\text{source}} $ represents the strong motion records,
and 
$u_{\text{target}} $ represents the target motion 
to predict.

\begin{figure}
  \centering
  \includegraphics[width = 8cm]{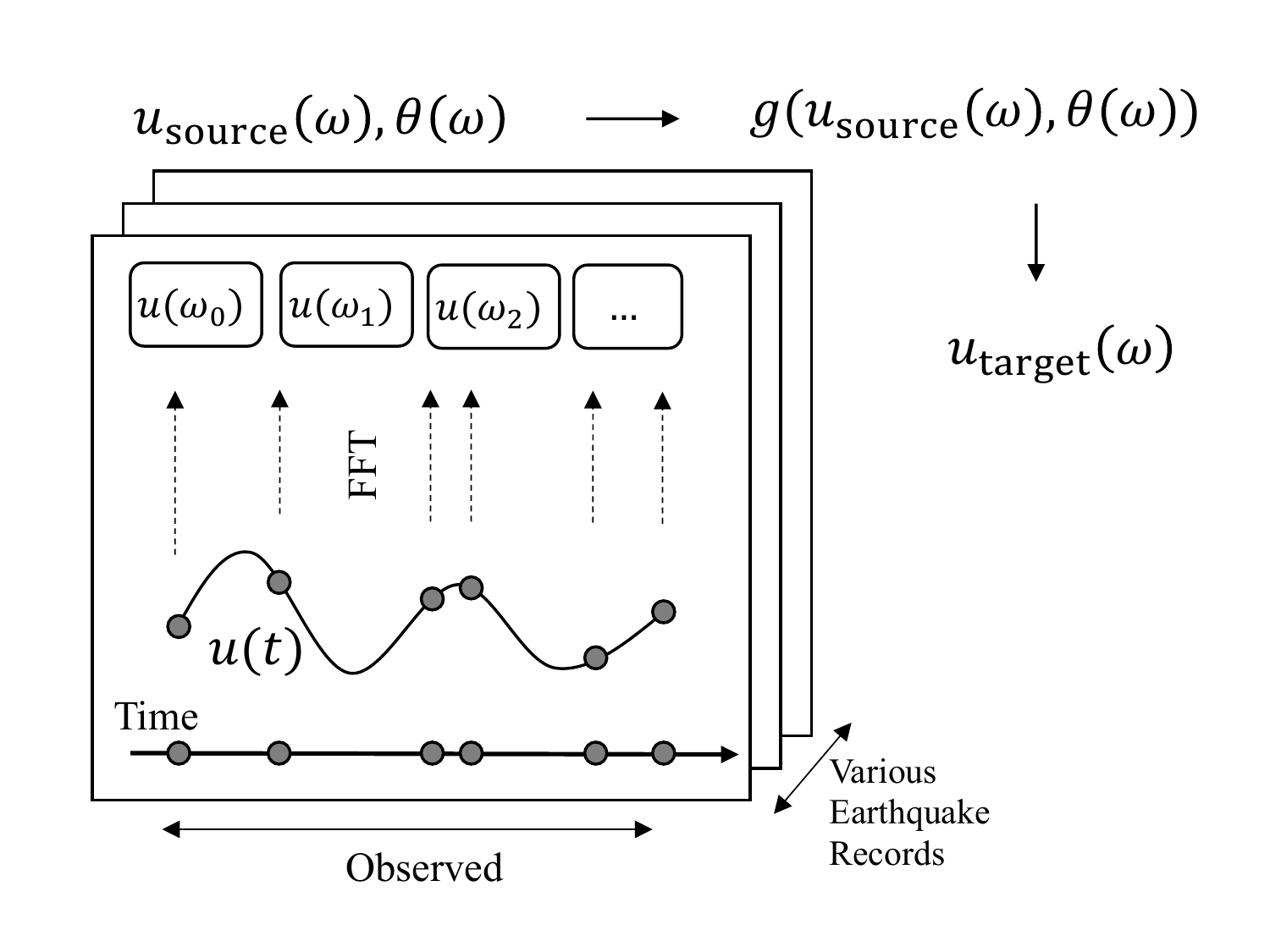}
  \caption{Computation graph for multiple degrees of freedom
  	using neural network.}
  \label{fig_mdof_computation_graph}
\end{figure}

The computation graph of MDOF system is illustrated 
in Figure~\ref{fig_mdof_computation_graph}.

Practically, 
in earthquake engineering, 
the acceleration at the top of the structure
is essential to the performance-based seismic design.
The dynamics response of floors are also important
to reduce the economic cost due to earthquakes.
Engineers can choose the specific targets to
estimate the dynamics responses
at different performance levels.
Compared to finite element simulation, 
a neural network allows engineers
to explicitly control the trade-off 
between accuracy and performance.
In most scenarios, with a few targets to predict,
neural network works 
much faster than 
a full scale finite element analysis.
Although the finite element method
is used to generate the ground truth
data in this work, 
neural networks can also work and be calibrated with 
experimental data directly without any modification.
%
%

\section{Application}\label{sec_application}

\begin{figure}
  \centering
  \includegraphics[width = 8cm]{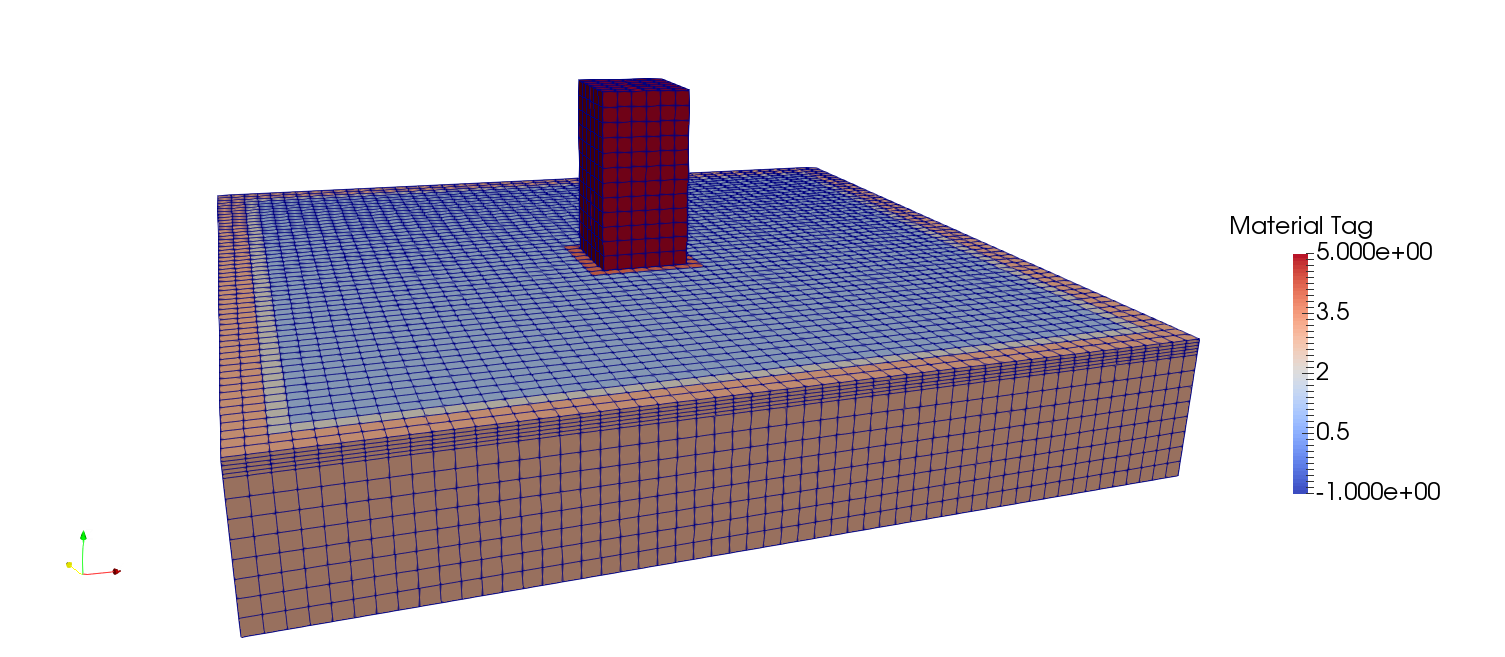}
  \includegraphics[width = 8cm]{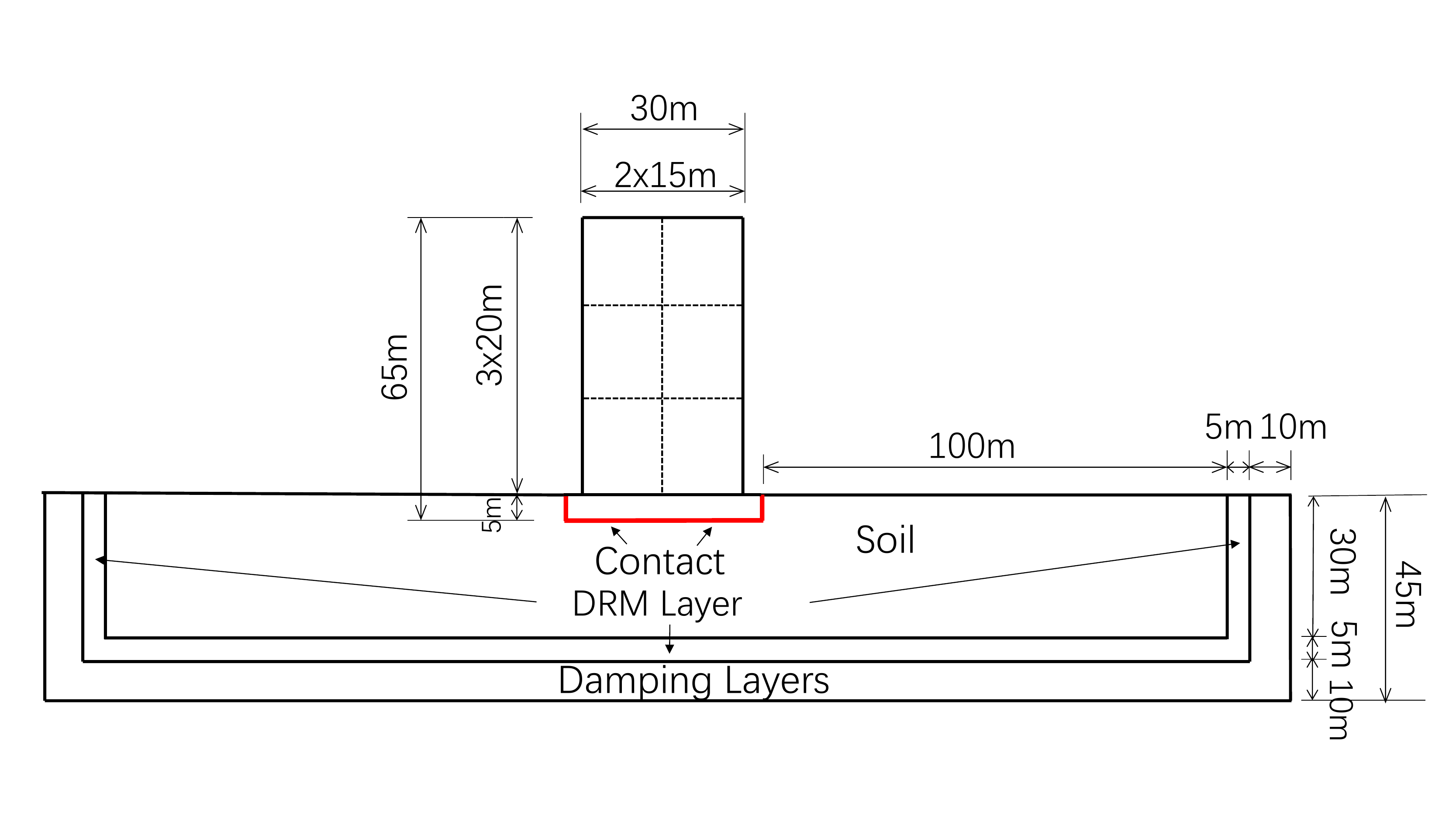}
  \caption{Materials types and geometries
  of the simulation model.}
  \label{fig_decon_1D_motion_3D_model}
\end{figure}

The application is 
an illustrative 3-story upper structure 
resides on top of the foundation and
soil layers.
The earthquake excitation is applied at the bottom
of the model.
The material types and geometries of 
the model are shown in 
Figure~\ref{fig_decon_1D_motion_3D_model}.

\textbf{Domain Reduction Methods}

Domain reduction method (DRM)\cite{bielak2003} is 
one of the best methods  
to apply 3-D seismic motions to 
seismic numerical models.
DRM has a two-stage strategy for the complex, realistic
3-D earthquake engineering simulation. 
The first stage is the generation of free field models 
with correct geology. 
And the second stage is the application of the 
generated free-field motion to the structure of interest. 
The DRM layer here is
modeled as a single layer of elastic soil. 
The damping layers adjacent to DRM layer were
modeled to prevent the reflected waves. 

\textbf{Structure Model}

The upper structure consists of a 3-story building
and a shallow foundation. 
The 3-story building consists of 2 floors of 
0.3 m thickness,
ceiling floor of 0.5 m thickness,
the exterior wall of 0.8 m thickness,
and interior walls of 0.2 m thickness.
The exterior and interior walls 
are embedded down to the depth of the foundation.
The foundation is a square shallow footing of size
 160 m$^2$ and thickness of 5.0 m.
The upper building is modeled as shell elements,
and foundation as linear brick elements,
both having the properties of concrete of 
elastic Young's modulus 30 GPa, Poisson ratio
0.22 and density 2100 kg/m$^3$.
The upper structure has a natural frequency 
of 3.47 Hz.

\textbf{Soil Model}

\begin{figure}
  \centering
  \includegraphics[width = 6cm]{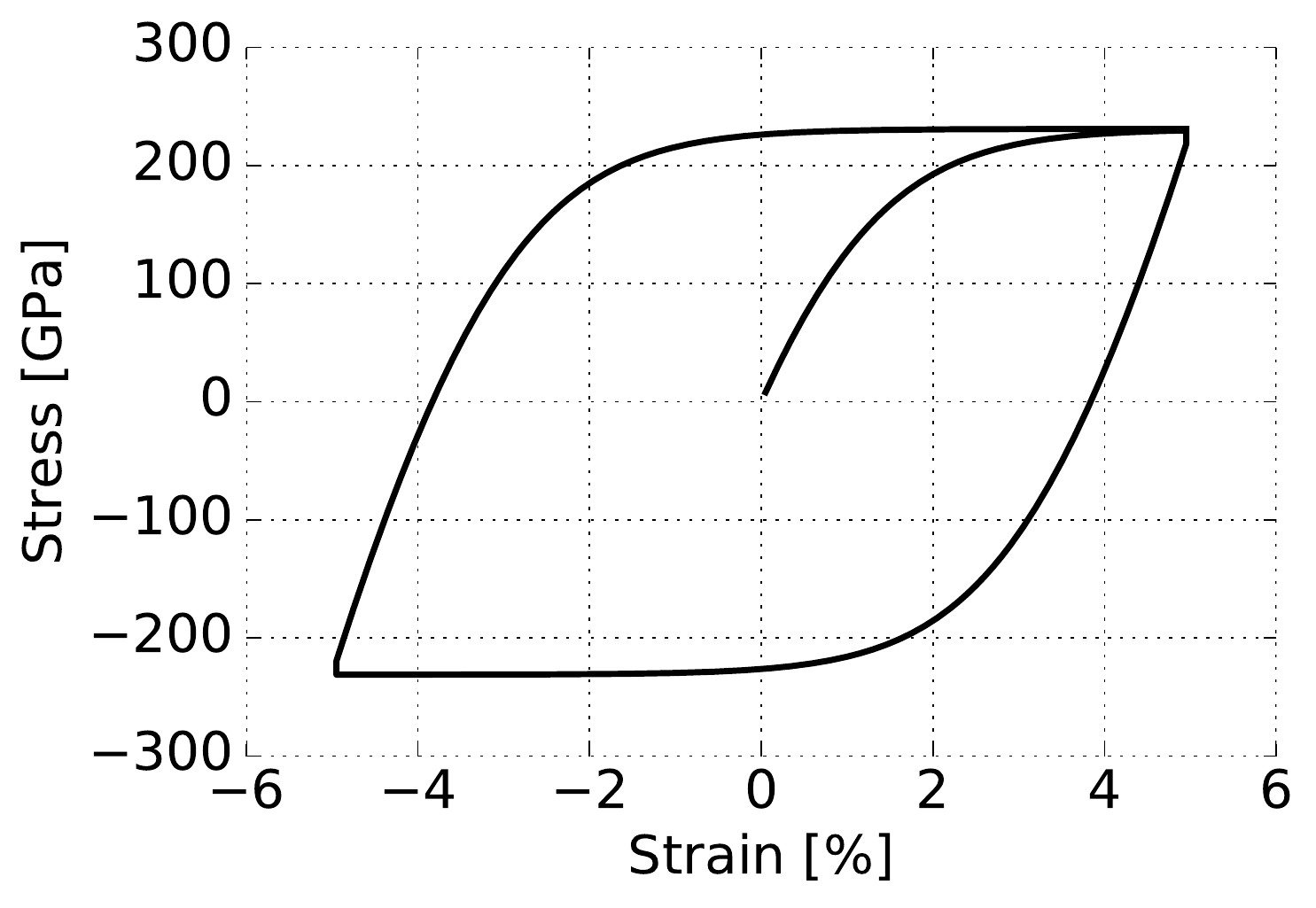}
  \caption{Soil modeling: stress-strain response curve}
  \label{fig_soil_modeling}
\end{figure}

The depth of the soil modeled below the foundation 
was 30 m, which is also the depth of 
DRM layer.
It is assumed that within this range 
the soil will have plasticity because
of its self-weight, structure, and seismic motions.
The soil is assumed to be a stiff saturated-clay
with undrained behavior having a shear velocity
of 450 m/s, unit weight of 20 kPa,
and the Poisson ratio of 0.15. 
The soil material model applies
the nonlinear kinematic hardening of 
Armstrong Frederick~\cite{jiang1996characteristics} type. 
The nonlinear inelastic model was calibrated
for yield strength achieved at 0.01 $\%$
shear strain.
The stress-strain response for the nonlinear
material model is shown in 
Figure~\ref{fig_soil_modeling}.

\textbf{Energy Dissipation}

In this MDOF application, 
the energy dissipation in decoupled elastic plastic material 
under isothermal condition~\cite{yang2018,yang2019energy} is 
given by
\begin{equation}
\Phi = \sigma_{ij} \dot{\epsilon}_{ij} - \sigma_{ij} \dot{\epsilon}_{ij}^{el} - \rho \dot{\psi}_{pl}
\ge 0
\label{eq_internal_dissipation_final}
\end{equation}

where $\Phi$ is the change rate  of energy dissipation per unit volume, 
$\sigma_{ij}$ and $\epsilon_{ij}$ are 
the stress and strain tensors respectively, 
$\epsilon_{ij}^{el}$ is the elastic part of the strain tensor, $\rho$ is the mass density of the material, and $\psi_{pl}$ 
is the plastic free energy per unit volume. 
Note that Equation~\ref{eq_internal_dissipation_final} is derived 
from the first and second laws of thermodynamics, 
which represents the conditions of energy balance and 
nonnegative rate of energy dissipation, 
respectively~\cite{yang2020plastic,yang2019energy2}. 
Considering all possible forms of energy, 
the energy balance between input mechanical work $W_{\text{Input}}$ and 
the combination of internal energy storage $E_{\text{Stored}}$ and 
energy dissipation $E_{\text{Dissipated}}$ can be expressed as
\begin{equation}
W_{\text{Input}} = E_{\text{Stored}} + E_{\text{Dissipated}} = E_{\text {KE}} + E_{\text {SE}} + E_{\text {PF}} + E_{\text {PD}}
\label{equation_energy_balance}
\end{equation}
where $E_{\text{KE}}$ is the kinetic energy, 
$E_{\text{SE}}$ is the elastic strain energy, 
$E_{\text{PF}}$ is the plastic-free-energy, 
and $E_{\text{PD}}$ is the energy dissipation due to material plasticity.
Note that the plastic dissipation term $E_{\text{PD}}$ includes energy dissipated in both elastic plastic solids (soil) and contact elements.

\textbf{Simulation and Results}

\begin{figure}
  \centering
  \includegraphics[width = 8cm]{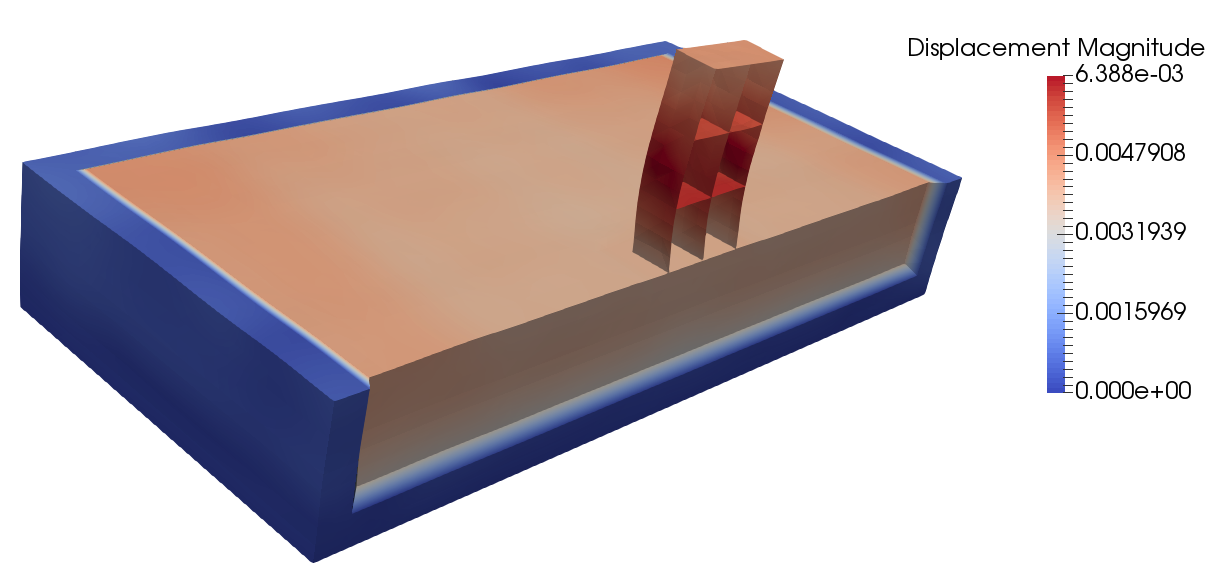}
  \caption{Illustrative deformation results
  of the simulation model at a time step.}
  \label{fig_decon_3D_motion_3D_model_results_structure}
\end{figure}

The illustration results of the simulation at
a time step are shown in Figure
~\ref{fig_decon_3D_motion_3D_model_results_structure}.
For this application, 
the neural network is trained using
earthquake records from PEER database~\cite{chiou2008}.
The testing data are untrained new earthquake records.
The training data consist of 100 earthquake records
from PEER database and each record has around 2000 timesteps.
In this work, the total number of earthquake timesteps 
used in training is around 225,000. 
The test data consist of 5 untrained earthquake records
with around 10,000 timesteps in total.
The prediction target is the dynamics response at the top
of the structure,
which plays an important role in the performance-based 
seismic design.

\begin{figure}
  \centering
  \includegraphics[width = 9cm]{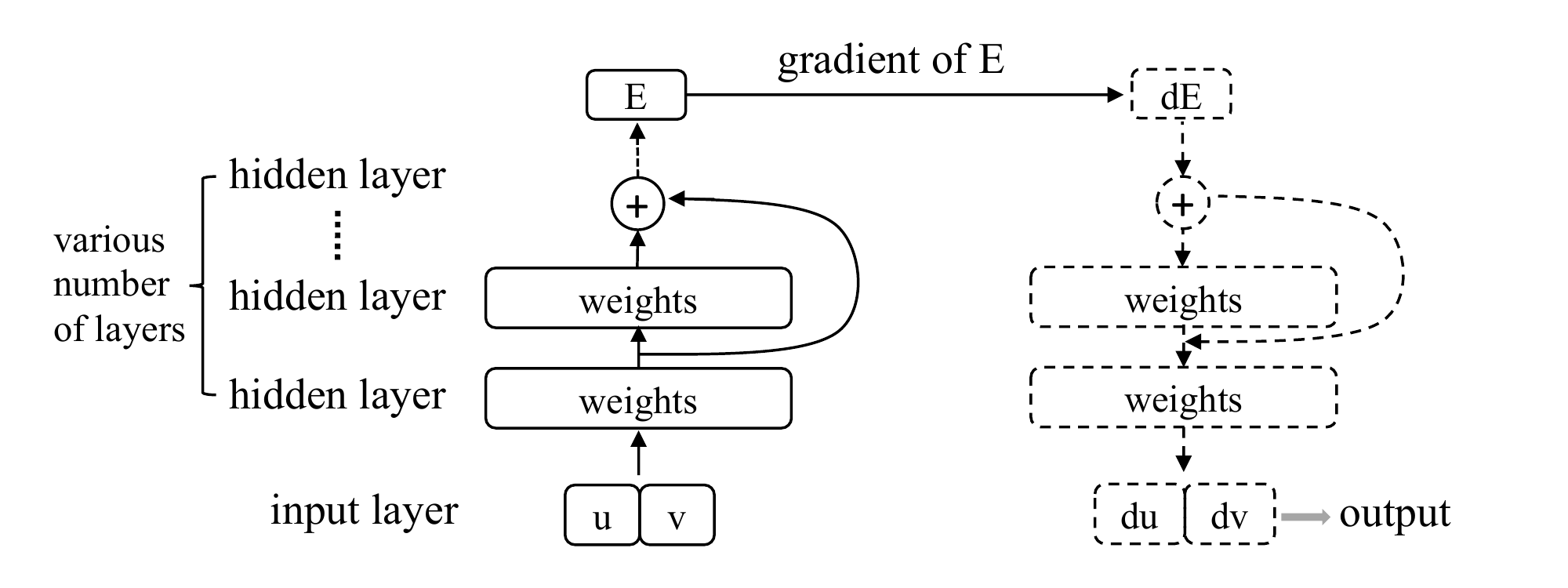}
  \caption{Illustrative architecture of the
  proposed energy conservation neural network.}
  \label{fig_ecnn_arch_various}
\end{figure}

The detailed parameter settings of the neural network
is as follows.
Firstly, the input layer transforms the number of
degrees of freedom to the number of hidden dimensions.
Secondly, multiple fully connected
hidden layers process the input data
consecutively. In addition, 
identity shortcut connections are added that
skips one or more hidden layers. 
Finally, the output layer transforms the data
to a single scalar, which represents the system energy
of the building structure.
The architecture of the energy-conservation 
is visualized in Figure~\ref{fig_ecnn_arch_various}.
Besides, orthogonal initialization is used 
as the initial weights.
Across each layer, ReLU (rectified linear 
unit)~\cite{hahnloser2000} is used as the activation function.
During the back-propagation, Adam is used as the optimizer 
with learning rate at 10$^{-3}$ and weight decay at 10$^{-4}$.
To demonstrate the efficacy of the proposed model,
the parameter settings and 
performance are extensively compared
with different number of hidden layers
and dimensions, 
as listed in 
Table~\ref{table_different_dimensions}
and Table~\ref{table_different_layers}.
\begin{table}[!htbp]
\centering
\begin{tabular}{ccccc}
\toprule
 \# Layers & \# Dimensions & \# Params & Baseline Error & ECNN Error \\ \hline
 2 &  50 & 5k & 2.77e-4 & 9.45e-5 \\ \hline
 2 & 200 & 82k & 2.22e-4 & 1.03e-4 \\ \hline
 2 & 350 & 248k & 1.13e-4 & 5.31e-5 \\
\toprule
\end{tabular}
\caption{The number of model parameters and test errors with different dimensions.}
\label{table_different_dimensions}
\end{table}
\begin{table}[!htbp]
\centering
\begin{tabular}{ccccc}
\toprule
 \# Layers & \# Dimensions & \# Params & Baseline Error & ECNN Error \\ \hline
2 & 200 & 82k & 2.22e-4 & 1.03e-4 \\ \hline
4 & 200 & 162k & 1.18e-4 & 3.15e-5 \\ \hline
6 & 200 & 243k & 1.07e-4 & 2.41e-5 \\
\toprule
\end{tabular}
\caption{The number of model parameters and test errors with different number of hidden layers.}
\label{table_different_layers}
\end{table}

\begin{figure}
  \centering
  \includegraphics[width = 7cm]{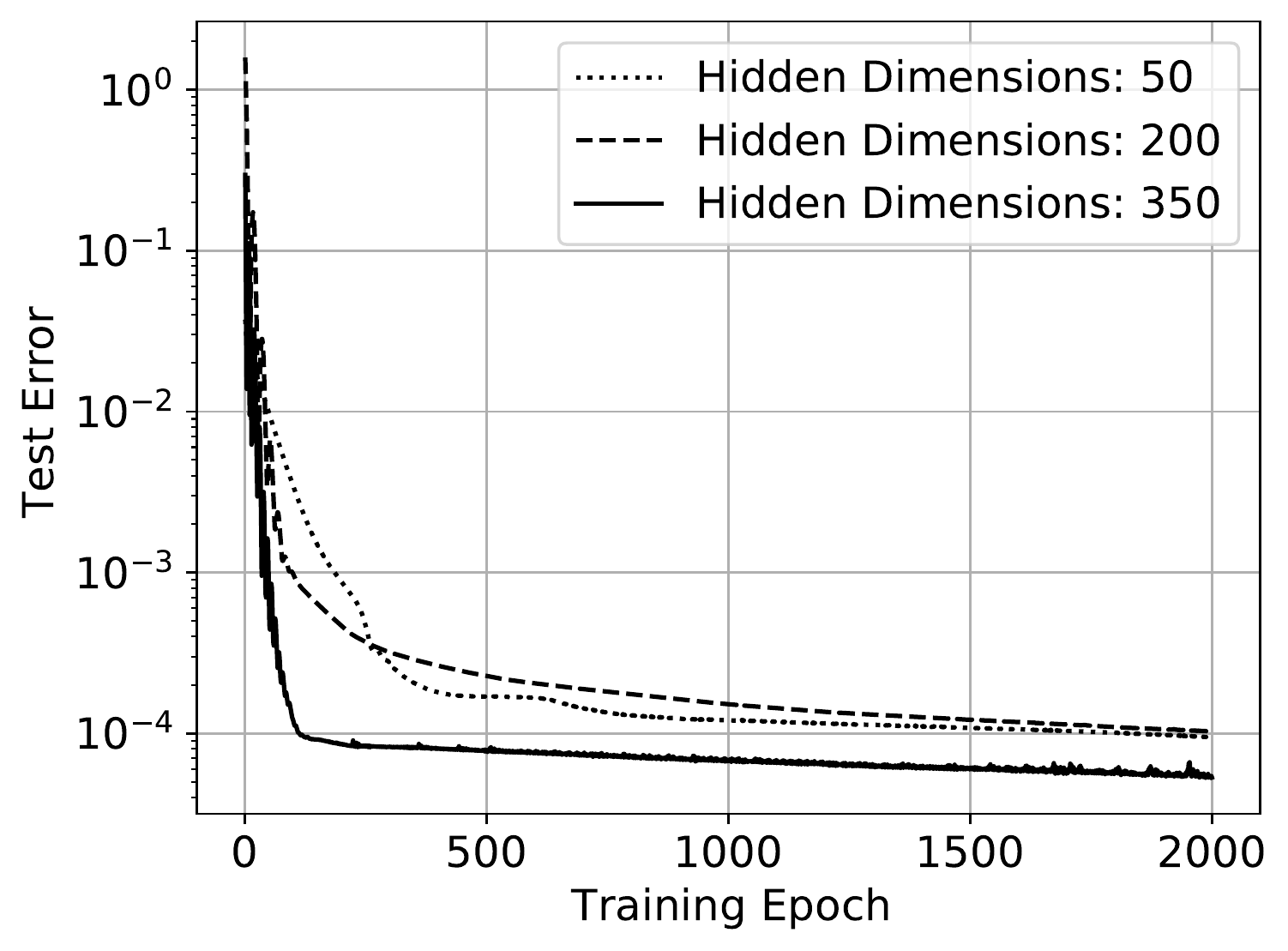}
  \caption{Test error over the entire training set 
    at the end of each epoch. 
    Wider dimensions tend to have smaller
    testing error.}
  \label{fig_stats_by_dim}
\end{figure}

\begin{figure}
  \centering
  \includegraphics[width = 7cm]{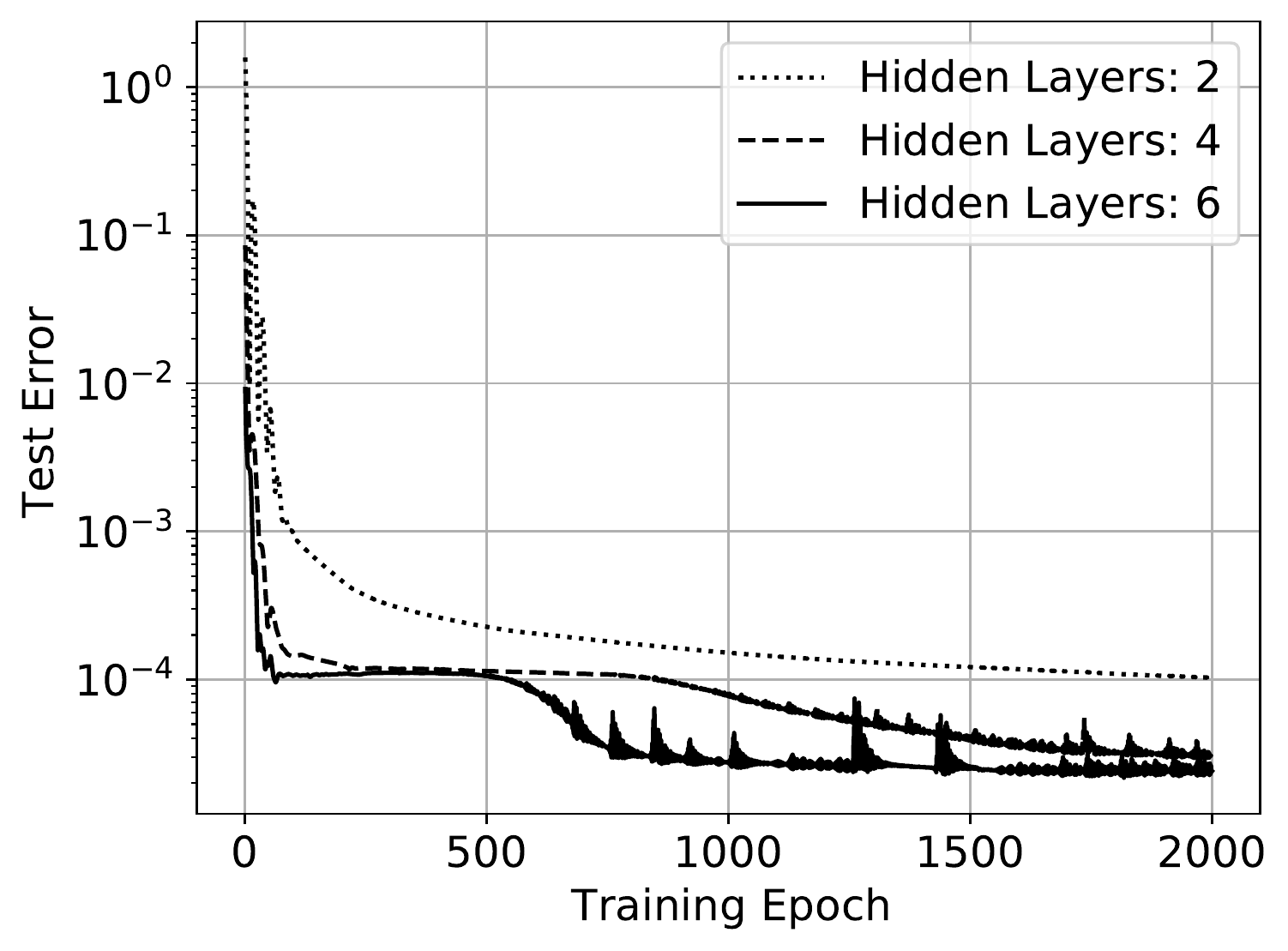}
  \caption{Test error over the entire training set 
    at the end of each epoch. 
    Testing error decreases with the increasing 
    number of hidden layers.}
  \label{fig_stats_by_layer}
\end{figure}

\begin{figure}
  \centering
  \includegraphics[width = 7cm]{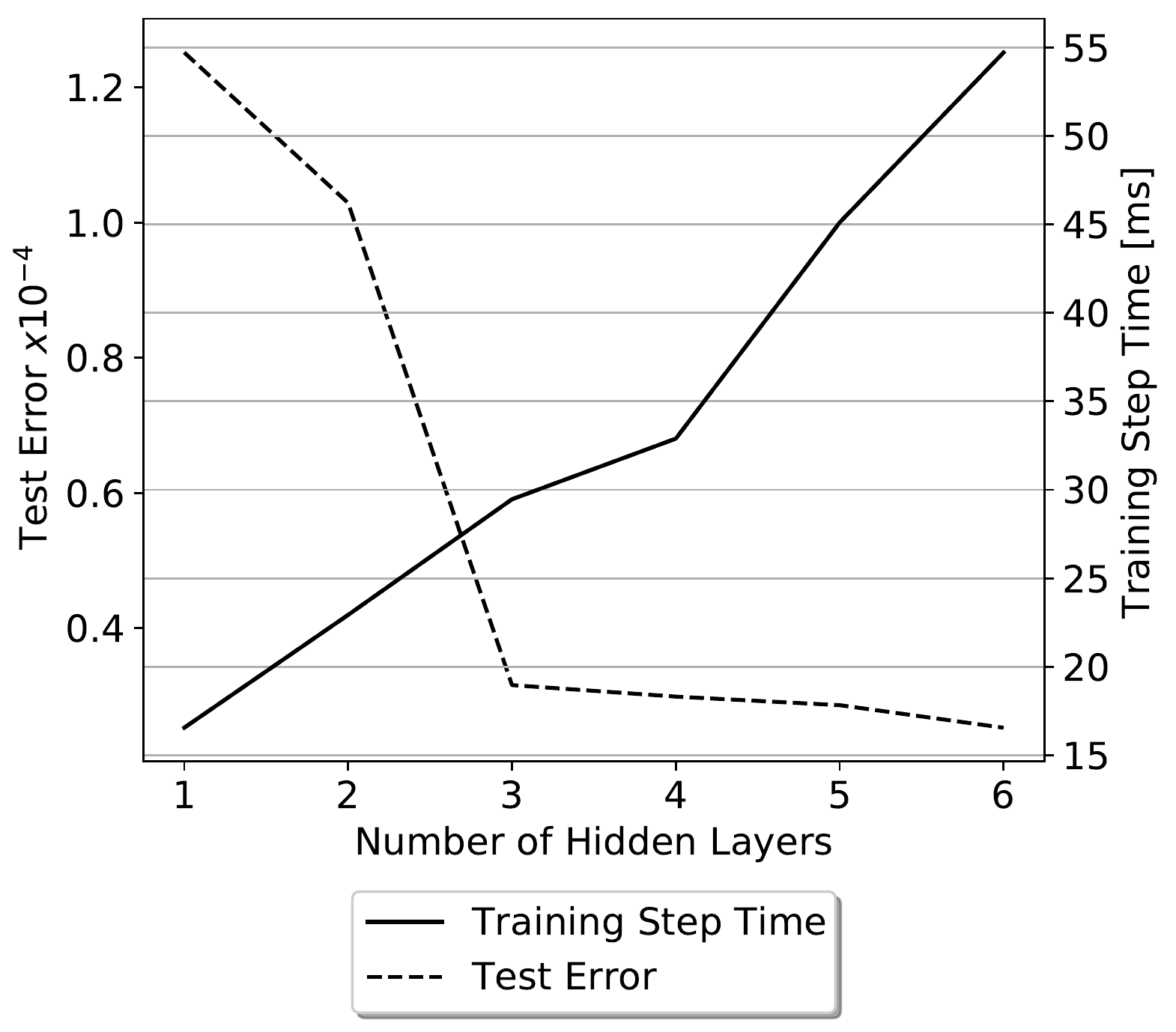}
  \caption{The trade-off between accuracy and speed with 
    increasing model depth. 
    Lower is better for both axes.}
  \label{fig_stats_layer_time}
\end{figure}

Furthermore, as shown in Figure~\ref{fig_stats_by_dim}
and Figure~\ref{fig_stats_by_layer},
the training error and test error decrease
with the increasing number of dimensions 
and layers. 
The increased accuracy is at the cost 
of a longer training time,
illustrated 
in Figure~\ref{fig_stats_layer_time}.

%
%

\section{Discussion}\label{sec_discussion}

Energy conservation neural networks (ECNN) were 
proposed in this work. 
Instead of training on the structural dynamics output
directly, 
ECNN uses system energy as the last layer 
of the neural network.
The gradient of the system energy is leveraged 
to calculate the structure states in the next step.
The proposed model takes advantage of the 
underlying automatic differentiation graph directly 
without adding extra parameters. 
While conventional neural networks describes the 
real world from data,
ECNN incorporates physics principles 
to the neural network.
The resulting method showcase a series 
of promising results from a fundamental
single-degree-of-freedom system to
a complicated multiple-degrees-of-freedom system.
Statistically, the physics rule is engraved
into the neural network and
the solution space of the neural network
is reduced.
The proposed model opens the path for
limiting the solution space of neural network
in a physically feasible
world 
and demonstrates
the possibility to endow machine learning
with the powerful capacity of mathematical physics rules
in the real-world.
As machine learning technology is continuing 
to grow fast both in terms of 
methodological development and hardware capability,
the proposed model can benefit practitioners 
across a wide range of scientific domains.
However, the proposed model should not be viewed
as replacements of the classical
numerical methods for solving structural dynamics problems.
Such methods have matured over years, which have met the robustness and 
efficiency requirements in practice.
Nevertheless, this work favors implementation 
simplicity and the rapid development and testing
of new ideas,
potentially opening
the path for the data-driven computation 
of structural dynamics problems.

\section{Scope and Limitations}\label{sec_scope_and_limitations}

\textbf{Uniqueness}

Since the structural dynamics problem can be abstracted into an 
initial value problem, the uniqueness of the structural
dynamics problem is equivalent to the uniqueness of 
the solution to an initial value problem.
According to Picard's existence 
theorem \cite{coddington1955}, 
the solution is unique if the differential equation is 
uniformly Lipschitz continuous. 
This theorem holds for the energy-conservation neural network, 
which has finite weights and uses Lipshitz nonlinearities 
with ReLU 
as the activation function.

\textbf{Time-series prediction}

In the SDOF system, the time-series prediction works well,
as shown in Figure~\ref{fig_sdof_constant_comparison}
and Figure~\ref{fig_sdof_damping_comparison}.
In the MDOF system, the acceleration and 
displacement amplitude is predicted in the 
frequency domain, 
which are the essential parameters for structural design
in earthquake engineering.
A proper prediction of the time-series in the MDOF system
can be achieved along with structural geometry.

\textbf{Error and tolerance}

Keeping the error within tolerance 
is crucial to structural dynamics in earthquake engineering.
In conventional numerical methods of structural dynamics,
the error comes from time-series discretization 
and geometry discretization.
So, conventionally, the error is controlled indirectly by time step-size
and mesh size.
With the neural network in this work,
the error is controlled directly during the training stage.
While training error evaluates how 
well the neural network matches the training data,
testing error shows if the pattern learned by the neural network
fits new data.
Besides the direct control over the training error and tolerance, 
the neural network allows the user to trade off speed for precision,
which means that 
the user can choose a good enough error tolerance 
for the best performance.

\section{Conclusion and Future Work}\label{sec_conclusion_and_future_work}

The classic and energy-conservation neural networks
are investigated to predict the structural dynamics response
for both SDOF and MDOF systems.
With the learned physics laws, 
the predicted structural dynamics response 
is more accurate.
Besides accuracy, the proposed neural network 
predicts the derivative of the structural states
instead of the discrete sequence of structural states.
With the predicted structural states derivatives, 
the actual structural response 
can be integrated with arbitrary time step-size,
which provides the time-continuous solution.
Furthermore, unlike the discretization errors 
in the conventional numerical methods for structural dynamics,
the proposed neural network model 
allows the explicit control of the trade-off between
computation speed and accuracy.
In addition, for MDOF systems,
the acceleration and displacement are predicted in the frequency domain.

For future works, although
the predicted frequency-domain results in the MDOF system
are essential for structural design,
a time-domain prediction in the MDOF system
is able to give more detailed information
for the structural dynamics design.
The time-domain prediction in the MDOF system
can be achieved with structural geometry.
In addition, the proposed approach is a new method 
in structural dynamics analysis.
There are many opportunities to 
improve the performance and to solve specific application
problems.
On the performance side, 
dimensionality reduction or reduced order modeling is 
a popular technique 
in both the machine learning and structural dynamics
community.
Using reduced order modeling to accelerate the 
simulation is worth further exploration.
On the application side,
performance-based building design is an new engineering
approach to design elements of a building upon 
performance goals and objectives.
The proposed method in this paper can be 
naturally customized to 
fit various performance indicators as needed.
Besides, from the machine learning perspective,
the possibility of transfer learning
on different building structures is also a promising 
opportunity for future study.

Last but not least, although the training data
for the neural network
in this paper are from finite element simulation,
training data from the realistic experiments will work as well
without any modification to the neural network.

\section*{Acknowledgments}

The earthquake records used in this paper are from
the PEER NGA online ground motion database.
We acknowledge contributions of a significant number of PEER researchers, including junior and senior researchers, post-doctoral fellows, graduate and undergraduate students, and practicing earthquake engineers and scientists.

This research did not receive any specific grant from funding agencies in the public, commercial, or not-for-profit sectors.


\bibliography{mybibfile}

\end{document}